\documentclass[%
 reprint,
superscriptaddress,
preprintnumbers,
nofootinbib,
 amsmath,amssymb,
 aps, prd,
twocolumn,
]{revtex4-1}

\usepackage{graphicx}
\usepackage{dcolumn}
\usepackage{bm}
\usepackage[colorlinks=true, linkcolor=blue, citecolor=blue, urlcolor=blue]{hyperref}
\usepackage{lipsum}
\usepackage[usenames,dvipsnames]{xcolor}
\usepackage{soul}

\usepackage{fontawesome}

\begin{document}


\title{Probing superheavy dark matter through
lunar radio observations of ultrahigh-energy neutrinos and the impacts of neutrino cascades}

\author{Saikat Das}
\email{saikat.das@yukawa.kyoto-u.ac.jp}
\affiliation{Center for Gravitational Physics and Quantum Information, Yukawa Institute for Theoretical Physics, Kyoto University, Kyoto 606-8502, Japan}

\author{Jose Alonso Carpio}
\email{jose.carpiodumler@unlv.edu}
\affiliation{Department of Physics \& Astronomy; Nevada Center for Astrophysics, University of Nevada, Las Vegas, NV 89154, USA}

\author{Kohta Murase}
\email{murase@psu.edu}
\affiliation{Department of Physics; Department of Astronomy \& Astrophysics; Center for Multimessenger Astrophysics, Institute for Gravitation and the Cosmos, The Pennsylvania State University, University Park, Pennsylvania 16802, USA}
\affiliation{Center for Gravitational Physics and Quantum Information, Yukawa Institute for Theoretical Physics, Kyoto University, Kyoto 606-8502, Japan}




\date{\today}
             

\begin{abstract}
Ultrahigh-energy neutrinos (UHE$\nu$s) can be used as a valuable probe of superheavy dark matter above $\sim 10^9$~GeV, the latter being difficult to probe with collider and direct detection experiments due to the feebly interacting nature. Searching for radio emissions originating from the interaction of UHE$\nu$s with the lunar regolith enables us to explore energies beyond $10^{12}$ GeV, which astrophysical accelerators cannot achieve. Taking into account the interaction of UHE$\nu$s with the cosmic neutrino background and resulting standard neutrino cascades to calculate the neutrino flux on Earth, for the first time, we investigate sensitivities of such lunar radio observations to very heavy dark matter. We also examine the impacts of cosmogenic neutrinos that have the astrophysical origin. We show that the proposed ultra-long wavelength lunar radio telescope, as well as the existing low-frequency array, can provide the most stringent constraints on decaying or annihilating superheavy dark matter with masses at $\gtrsim 10^{12}$~GeV. The limits are complementary to or even stronger than those from other UHE$\nu$ detectors, such as the IceCube-Gen2 radio array and GRAND.
\end{abstract}

\maketitle



\section{\label{sec:intro}Introduction\protect}
Very heavy dark matter (VHDM) at energies $m_{\rm DM}c^2\gtrsim 1$~TeV may be out of reach of the current collider experiments due to their high masses. Direct search is also challenging owing to low rates in nuclear scattering experiments. Searching for indirect signatures of their decay and annihilation, producing high-energy cosmic rays, neutrinos, and $\gamma$ rays, are particularly important (e.g., Refs.~\cite{Bertone:2004pz, Klasen:2015uma, Arguelles:2019xgp}). VHDM can be produced either nonthermally or thermally in the early Universe \citep{Greene:1997ge, Chung:1998bt, Chung:1998zb, Kuzmin:1998uv, Chung:1998ua, Chung:1998rq, Chung:1999ve, Kuzmin:1999zk, Chung:2001cb, Kolb:2007vd,  Kannike:2016jfs, Kim:2019udq, Kramer:2020sbb, Ling:2021zlj}. Their mass can be larger than the unitarity limit \citep{Griest_1990} and may extend up to the Grand Unification Theory (GUT) energy scale at $\Lambda_{\rm GUT}\sim 10^{15}-10^{16}$~GeV and even beyond, going up to as high as the Planck energy scale at $\sim10^{19}$~GeV \citep{Garny:2015sjg, Hooper:2019gtx, Mambrini:2021zpp}. VHDM should have a lifetime longer than the age of the Universe but could decay into Standard Model (SM) particles, including quarks, leptons, and vector/scalar bosons. Currently operating and multimessenger detectors in the TeV-EeV energy range have the potential to detect SM particles from VHDM decay or annihilation, viz., $e^\pm$, $\nu$, $\overline{\nu}$, $p$, $\overline p$, and $\gamma$ rays \citep[e.g.,][]{Berezinsky:1997hy, Kuzmin:1997jua, Aloisio:2006yi, Kachelriess:2007aj, Murase:2012xs, Esmaili:2012us, Murase:2015gea, Kachelriess:2018rty, Ishiwata:2019aet, Guepin:2021ljb, Arguelles:2022nbl, Fiorillo:2023clw, Munbodh:2024ast}. 

An extremely large exposure is required to detect ultrahigh-energy cosmic rays (UHECRs; $\gtrsim3\times10^{9}$~GeV) and ultrahigh-energy neutrinos (UHE$\nu$s), owing to their low event rates \citep{PierreAuger:2020qqz, PierreAuger:2021hun, PierreAuger:2019fdm, Abbasi:2021hk, Abbasi:2021hO}. 
The photomeson interaction process of UHECR protons with the cosmic microwave background (CMB) and the extragalactic background light (EBL) \citep{Greisen:1966jv, Zatsepin:1966jv}, leads to the production of ``cosmogenic'' neutrinos \citep{Berezinsky:1969erk, Yoshida:1993pt, Engel:2001hd, Berezinsky:2002nc, Ave:2004uj, Seckel:2005cm, Hooper:2004jc, Stanev_2006, Takami:2007pp, Kotera_2010, Ahlers:2010fw, Fang:2017zjf, AlvesBatista:2018zui, Das:2018ymz, Heinze:2019jou, Zhang:2018agl, Das:2020nvx, Jiang:2020arb}. However, the dominance of heavier nuclei (such as $^{28}$Si or $^{56}$Fe) at the highest energies up to a few times $10^{20}$~eV is inferred from recent observations, thus reducing the flux of protons for a rigidity-dependent injection spectrum \citep{PierreAuger:2010ymv, PierreAuger:2014gko, PierreAuger:2016use}. 
Sources of UHECR protons should be more efficient neutrino emitters than those of UHECR nuclei if the sources allow the survival of nuclei \citep{Murase:2010gj}. 
Source neutrinos could potentially achieve extremely high energies, but their spectrum is unlikely to extend beyond $\sim10^{12}-10^{13}$~GeV \cite{Thompson:2011xb}. For sources contributing to the UHECR flux on Earth, the neutrino spectrum is likely to fall off sharply beyond a few times $10^9$~GeV \citep[e.g.,][]{Kotera_2010, Fang:2017zjf, Zhang:2018agl, AlvesBatista:2018zui, Muzio:2023skc}. 

Detecting neutrinos beyond the energy range of astrophysical neutrinos possesses the potential to unveil physics beyond the Standard Model (BSM). VHDM decay can dominantly contribute to the neutrino flux in this regime. Such high-energy frontiers are interesting in the context of neutrino detection. UHE$\nu$s can interact with a dense dielectric medium, initiating particle showers. The resultant charge asymmetry in the shower front can produce a short, coherent burst of radio emission. Particle detectors, sensitive to radio photons from such neutrino interactions, have been proposed and employed \citep{Ackermann:2022rqc, Ackermann:2019ows, Ackermann:2019cxh}. The Moon can be projected as a natural detector with a large collector area. UHE$\nu$s interacting with the lunar regolith can produce Cherenkov radiation via the Askaryan effect, which can be detected using a sensitive radio telescope \citep{Hankins_1996, Hankins:2000fg, Gorham:2003da, Stal:2006te, Burns:2011wf}. 

The aperture of neutrino detection using the Moon can be larger than detectors for radio emission from neutrino-initiated cascades in ice \citep[e.g., IceCube-Gen2 radio array][]{IceCube:2019pna, IceCube-Gen2:2020qha}) due to a greater detector volume achievable within the field of view of a radio telescope. The technique was applied to the PARKES radio telescope, and later, observational limits were improved using the PARKES-ATCA joint observations \citep{Hankins_1996, James:2009rc, lunaska_2010}. The NuMoon experiment using observations by the LOFAR detector has placed upper limits on the UHE neutrino detection above $10^{13}$ GeV \citep{Krampah:2023mpj}. The LUNASKA experiment (Lunar ultrahigh-energy Neutrino Astrophysics with the SKA) will perform a lunar neutrino experiment with the  radio telescope Square Kilometer Array \citep{Bray_2015}. The proposed lunar Ultra-Long Wavelength (ULW) radio telescope has provided the projected all-flavor UHE$\nu$ flux sensitivity at energies higher than a few times $10^{12}$~GeV for different orbital altitudes of the radio antenna and assuming an exposure time of one year \citep{Chen:2023mau}. 

On the other hand, at extremely high energies, neutrinos may not travel from the sources to the Earth without any interaction. In SM, neutrinos originating from high redshifts can interact with the cosmic neutrino background (C$\nu$B) via neutrino-neutrino interactions \citep{Weiler_1982, Weiler_1984, YOSHIDA1994187, Yoshida:1996ie,Fodor:2002qf}. Beyond $10^{11}$~GeV, the Universe can be probed using neutrinos up to a ``horizon'' of $z\sim140$ \citep{Berezinsky_1992}. The resonant absorption due to neutrino-antineutrino annihilation ($\nu+\overline{\nu}\rightarrow$ anything) via the $Z$ boson has been studied earlier in detail and extended to include the thermal effect on the C$\nu$B \citep{DOlivo:2005edp, Barenboim:2004di}. Effects of warm or hot neutrinos are relevant for neutrino sources at $z\gtrsim16$ \citep{Lunardini:2013iwa}. Even applications to high-energy neutrino interactions beyond the Standard Model have been discussed \citep{Ioka:2014kca, Ng:2014pca, Shoemaker:2015qul,Murase:2019xqi,Carpio:2021jhu,Carpio:2022lqk}. Thus, this flux suppression is important to constrain the VHDM decay timescale or annihilation cross section. In this work, neutrino annihilation and scattering cross sections are treated at tree level. In addition to the flux suppression effects, which is considered in previous works on neutrino propagation, we also include the contributions from neutrino cascades. At very high redshifts $z\gtrsim 10$, reinjection of neutrinos at lower energies can impact the observed fluxes on Earth.

The differential sensitivity, $E^2\Phi$ of the current UHE$\nu$ detectors, including IceCube and Auger, is the best around $10^9$~GeV and worsens at higher energies. 
In this work, we aim to derive limits to the decay lifetime and annihilation cross section of VHDM particles in the mass range of $m_{\rm DM}\approx 10^9 - 10^{16}$~GeV, using UHE$\nu$s. The future detection of cosmogenic neutrinos in this energy range by  UHE$\nu$ detectors such as the Giant Radio Array for Neutrino Detection (GRAND) \citep{GRAND:2018iaj} and IceCube-Gen2 radio array \citep{IceCube:2019pna, IceCube-Gen2:2020qha} seems optimistic. Their projected sensitivities give the most stringent bounds on VHDM up to $\lesssim 10^{12}$ GeV. We find that at energies $\gtrsim 10^{12}$~GeV, the detection potential of the ULW telescope provides new constraints on VHDM, which is orders of magnitude higher than the currently operating or planned next-generation detectors in this energy range. The absence of an astrophysical background at $\gtrsim10^{12}$~GeV leads to the most stringent limits.
 
We describe our model considerations on the neutrino propagation and neutrino flux from VHDM decay/annihilation in Sec.~\ref{sec:methods} and Sec.~\ref{sec:nuvhdm}, respectively. We present our results on VHDM constraints in Sec.~\ref{sec:results} and discuss the implications in Sec.~\ref{sec:discussions}. Further details on solving the neutrino transport equation for extragalactic propagation are presented in Appendix~\ref{appendix:implicitmethod}.
In this work, we use natural units to write down analytical expressions for particle physics processes, such as cross sections $\sigma$, differential cross sections $d\sigma/d\varepsilon$ and associated kinematic formulas. We also use CGS units to represent the values.

\section{\label{sec:methods}UHE$\nu$ Propagation\protect}
The propagation of astroparticles can be calculated by solving Boltzmann equations, which have been extensively studied for electromagnetic cascades. We develop a numerical module for the UHE$\nu$ propagation within the framework of \textsc{AMES} (Astrophysical Multimessenger Emission Simulator)\footnote{\url{https://github.com/pegasuskmurase/AMES-GRBAfterglow}}. In this section, we describe the method of calculations and refer to Sec.~\ref{appendix:implicitmethod} for more details. 

Let $\tilde{\mathcal{Q}}_{\nu_\alpha}(\varepsilon,t)$ be the number of neutrinos $\nu_\alpha$ injected per time, with energies (in the comoving frame) between $\varepsilon$ and $\varepsilon + d\varepsilon$ at a time $t$. 
The UHE$\nu$s propagate in the bath of C$\nu$B neutrinos that serve as target particles for weak interactions. We consider Dirac neutrinos when providing expressions such as cross sections, but our results remain unaltered if neutrinos are Majorana particles. 

Massive neutrinos oscillate as they propagate from a source to an observer. However, oscillations eventually become suppressed by the wave-packet decoherence once the wave packets of the mass eigenstates get separated due to the difference in their group velocities. The overlap of the wave packets takes place while the propagation length is shorter than the coherence length $L_{\rm coh} = 2\varepsilon^2\sigma_x/\Delta m^2$, where $\sigma_x$ is the wave packet size and $\Delta m^2$ is the squared mass difference between the mass eigenstates \cite{Kayser1981,Beuthe:2001rc,Akhmedov:2009rb}. 
For C$\nu$B neutrinos, we assume wave-packet decoherence has occurred, as done in other works (e.g., \cite{Eberle:2004ua,Lunardini:2013iwa}). 
For example, in the case of dark matter decay with the lifetime $\tau_{{\rm DM}}$, $\sigma_x\sim \tau_{\rm DM}$ \cite{Akhmedov:2009rb}, the long lifetimes $\tau_{\rm DM}\gtrsim10^{26}\;{\rm s}$ are such that $\sigma_x$ is much longer than the oscillation length \cite{Giunti:1997wq}. For dark matter annihilation, the size of the wave packet is $\sigma_x\sim  t_{\rm DM-DM}$, where $t_{\rm DM-DM}^{-1}=n_{\rm DM}\langle\sigma v\rangle_{\rm DM}$ is the annihilation rate. 
Hence, both C$\nu$B neutrinos and UHE$\nu$s can be treated as mass eigenstates, so the injection rate, $\tilde{\mathcal{Q}}_{\nu_\alpha}$, can be converted into the rates in the mass basis via
\begin{equation}
\tilde{\mathcal{Q}}_{\nu_i} = \sum_{\alpha=e,\mu,\tau} |U_{\alpha i}|^2 \tilde{\mathcal{Q}}_{\nu_\alpha},
\label{FlavorToMassConversion}
\end{equation}
where $U$ is the Pontecorvo-Maki-Nagagawa-Sakata (PMNS) matrix. The same relationship is applied to antineutrinos, substituting $U_{\alpha i}\to U^*_{\alpha i}$. For neutrino scatterings during the propagation, the initial and final states are assumed to be mass eigenstates (i.e., we treat neutrinos as fully mixed in flavor space). This assumption is valid in the regime where the interaction length is longer than the coherence and oscillation lengths, which will hold in our work, where most of the VHDM contributions to the neutrino flux occur at $z\lesssim 10$. At higher redshifts, the C$\nu$B density is so large that neutrinos may interact more than once, necessitating a more dedicated approach (see, e.g., Ref.~\cite{Ema:2014ufa}). 

For the notations, we label 4-momenta as $p$ and 3-momenta as $\mathbf{p}$. Neutrino flavor (mass) eigenstates are labeled as $\nu_\alpha$ ($\nu_i$). We use the best-fit neutrino oscillation parameters from {\tt NuFIT 2022} \cite{Esteban2020}. 
We also adopt the masses, $m_1=0.0218$~eV, $m_2=0.0235$~eV, and $m_3=0.0547$~eV, consistent with the neutrino oscillation values of $\Delta m_{ij}^2$ from oscillation data \cite{Esteban2020} and the cosmological bound of $\sum m_\nu<0.12$~eV \cite{Planck:2018vyg}.

Under SM interactions, neutrino decoupling from matter yields a thermal C$\nu$B with a temperature at redshift $z$, $T_{{\rm C}\nu{\rm B}}(z)=(1+z)T_0$, where $k_BT_0=1.697\times10^{-4}$~eV is the C$\nu$B temperature today, as used in Ref.~\cite{Lunardini:2013iwa}, and $k_B=8.6173\times 10^{-5}$~eV K$^{-1}$ is the Boltzmann constant. Each C$\nu$B mass eigenstate has a number density of 
$n_{\nu_j}^{{\rm C}\nu{\rm B}}(z) = n_{\bar\nu_j}^{{\rm C}\nu{\rm B}}(z) = 56(1+z)^3$~cm$^{-3}$ (where neutrinos and antineutrinos are distinguished) and is treated as unpolarized \cite{Long:2014zva}. 
Each C$\nu$B neutrino species $B$ follows the Fermi-Dirac distribution
\begin{equation}
dn_{B}^{{\rm C}\nu{\rm B}} = \frac{d^3\mathbf{p}_B}{(2\pi\hbar )^3}\frac{1}{e^{c|\mathbf{p_B}|/k_B T(z)}+1}.
\label{Fermi-DiracDistribution}
\end{equation}

Let $\tilde{N}(\varepsilon,t) = dN/d\varepsilon$ be the comoving number of neutrinos per unit energy at time $t$. The neutrino transport equation for a neutrino species $A$ is given by
\begin{align}
\frac{\partial \tilde{N}_{A}(\varepsilon,t)}{\partial t}& = \frac{\partial}{\partial\varepsilon} [H(z(t))\varepsilon\tilde{N}_{A}(\varepsilon,t)]  -\tilde{N}_{A}(\varepsilon,t)\mathcal{A}_{A}(\varepsilon,t)  \nonumber \\
& + \int_\varepsilon^\infty d\varepsilon' \tilde{N}_{A}(\varepsilon',t) \mathcal{B}_{A\to A}(\varepsilon,\varepsilon')\nonumber \\
& + \sum_{B\neq A}\int_\varepsilon^\infty d\varepsilon'\tilde{N}_{B}(\varepsilon',t)\mathcal{C}_{B\to A}(\varepsilon,\varepsilon')\nonumber \\
& +\tilde{\mathcal{Q}}_{A}(\varepsilon,t),
\label{TransportEquation_Time}
\end{align}
where $A,B\in \{\nu_1,\nu_2,\nu_3,\bar{\nu}_1,\bar{\nu}_2,\bar{\nu}_3\}$ and $z(t)$ is the redshift at time $t$. We point out that the initial time $t=0$ corresponds to the injection time of the source located at some redshift $z_s$, and the propagation is carried out until the time when the neutrino reaches $z=0$. Redshift and time are related by
\begin{equation}
t = \int_z^{z_s}\frac{dz^\prime}{(1+z^\prime)H(z^\prime)},
\end{equation}
where the Hubble parameter $H(z)$ is given by
\begin{equation}
H(z) = H_0\sqrt{\Omega_\Lambda + \Omega_M(1+z)^3},
\end{equation}
assuming $\Omega_\Lambda=0.685$, $\Omega_M=0.315$, and the Hubble constant $H_0 = 67.3$~km~s$^{-1}$~Mpc$^{-1}$ \cite{Planck:2018vyg}.

The first term on the right-hand side of Eq. \ref{TransportEquation_Time} is the adiabatic loss rate. 
$\mathcal{A}_A$ is the total interaction rate of $A$, while $\mathcal{B}_{A\to A}$ is the self-production rate of $A$ with energy $\varepsilon$ generated from the same particle type with energy $\varepsilon'$. $\mathcal{C}_{B\to A}$ is the production rate of particle type $A$ with energy $\varepsilon$ from other particle type $B$ with energy $\varepsilon'$. Note that the integrals above imply that the incident neutrino always loses energy, which is reasonable given that the C$\nu$B energy is negligible compared to $\varepsilon$ and $\varepsilon'$.

Within the SM, we consider the following tree-level processes in the calculation of UHE$\nu$ interactions with C$\nu$B neutrinos:
\begin{enumerate}
\item Annihilation to fermion pairs $\nu_i\bar{\nu}_i\to f\bar{f}$ mediated by the $Z$ boson in the $s$-channel
\item Neutrino-antineutrino scattering $\nu_i\bar{\nu}_j\to\nu_i\bar{\nu}_j$ mediated by the $Z$ boson in the $t$-channel
\item Neutrino-neutrino scattering $\nu_i\nu_j\to\nu_i\nu_j$ mediated by the $Z$ boson in the $t$-channel and, when $i=j$, in the $u$-channel as well.
\item Charged lepton production $\nu_i\nu_j\to l_i l_j$ and $\nu_i\bar{\nu}_j\to l_i \bar{l}_j$ mediated by the $W$ boson in the $t$-channel
\end{enumerate}

In our implementation, we only consider processes involving neutrinos in the final states for the particle reinjection rates $\mathcal{B}$ and $\mathcal{C}$. Hence, we do not reinject neutrinos from pion and muon decays produced during neutrino propagation (see e.g., Refs.~\cite{Li:2007ps,Esmaeili:2022cpz,Esmaeili:2023vyk}). However, all processes are considered for the absorption rate~$\mathcal{A}$. 

Among the aforementioned processes, the most relevant one is the $s$-channel resonant process, $\nu_i\bar{\nu}_i\to f\bar{f}$ annihilation, with a cross section given by \cite{PhysRevD.47.5247}
\begin{widetext}
\begin{equation}
\sigma (s)= \frac{2G_F^2m_Z^4}{3\pi}\frac{s}{(s-m_Z^2)^2+m_Z^2\Gamma_Z^2}\sum_f n_f\left[T_{3,f}^2-2T_{3,f}Q_f\sin^2\theta_w+2Q_f^2\sin^4\theta_w\right],
\label{schannelannihilation}
\end{equation}
\end{widetext}
where $G_F$ is the Fermi constant and $m_Z(\Gamma_Z)$ is the $Z$ boson mass (width). We define the Mandelstam variable \textbf{$s=2\varepsilon_{\rm I}\varepsilon_{\rm tar}(1-\beta_{\rm tar}\cos\theta)$} for the incident neutrino energy $\varepsilon_{\rm I}$, target neutrino energy $\varepsilon_{\rm tar}$,  target speed $\beta_{\rm tar}$, and angle $\theta$ between their momenta. The sum is done over fermions that satisfy $s>4m_f^2$, and {$Q_f(T_{3,f})$} is the fermion's electric charge (weak isospin), $n_f=1(3)$ for leptons (quarks), $\sin^2\theta_w\approx 0.23$, and $\theta_w$ is the Weinberg angle. 

The cross section has the maximum at $s=m_Z^2$. If the target neutrino is nonrelativistic ($m_{\rm tar}<T_{{\rm C}\nu{\rm B}}(z)$), which is typically the case at low redshifts, then the resonance is achieved if the incident neutrino has the energy
\begin{equation}
\varepsilon_{\rm I}^{\rm res} = \frac{m_Z^2}{2m_{\rm tar}} \simeq 4.2\times 10^{14}~{\rm GeV} \left(\frac{0.01 {\rm eV}}{m_{\rm tar}}\right).
\label{NR_Resonance}
\end{equation}

If $(1+z)\gtrsim m_{\rm tar}/\langle \sqrt{|\mathbf{p}|^2\rangle} \approx 16 (m_{\rm tar}/0.01\; {\rm eV})$, where $\langle |\mathbf{p}|^2\rangle$ is the thermal average of $|\mathbf{p}|^2$, neutrinos become relativistic and the resonance energy depends on the incident angle \cite{Lunardini:2013iwa}. We then have
\begin{equation}
    \varepsilon_{\rm I}^{\rm res} \simeq \frac{m_Z^2}{2(1+z)T_0(1-\cos\theta)} = \frac{2.4\times 10^{16}\;{\rm GeV}}{(1+z)(1-\cos\theta)}.
    \label{Rel_Resonance}
\end{equation}

When C$\nu$B neutrinos are considered at rest, Eq.~\ref{NR_Resonance} tells us that only three resonance energies exist, one for each of the different neutrino masses. However, relativistic C$\nu$B neutrinos allow for a broader range of resonance energies, depending on the angle between incident and target neutrinos. This leads to an effective cross section for collisions between neutrino species $A$ and $B$, 
\begin{equation}
\sigma_{{\rm eff},AB} = \dfrac{\displaystyle\int d^3\mathbf{p}_{B}\dfrac{dn_B^{{\rm C}\nu{\rm B}}}{
d^3\mathbf{p}_B}(1-\beta_B\cos\theta_{AB})\sigma_{AB}(s)}{\displaystyle\int d^3\mathbf{p}_{B}\dfrac{dn_B^{{\rm C}\nu{\rm B}}}{
d^3\mathbf{p}_B}},
\end{equation}
where $\beta_B$ is the speed of particle $B$ and $\theta_{AB}$ is the angle between the neutrino momenta.

\begin{figure}
\includegraphics[width=0.45\textwidth]{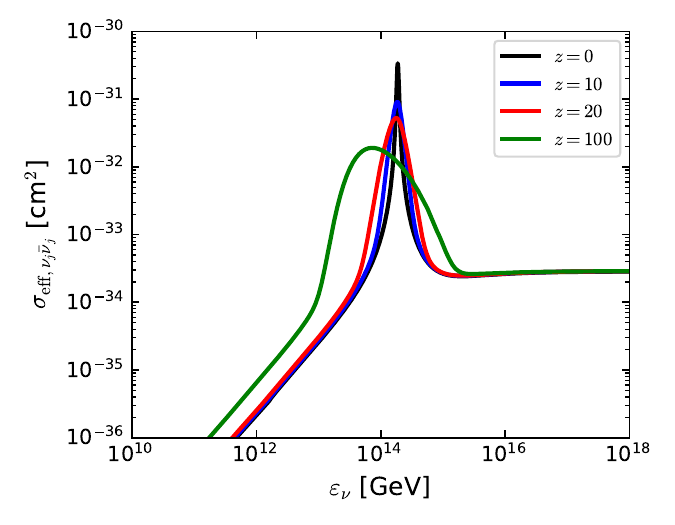}
\caption{Effective cross section, $\sigma_{{\rm eff},\nu_j\bar{\nu}_j}$, for a neutrino mass of $m_j = 0.0218\;{\rm eV}$. 
We show the cross section for C$\nu$B target neutrinos at $z=0$ (black), $z=10$ (blue), $z=20$ (red), and $z=100$ (green).}
\label{EffectiveCrossSection}
\end{figure}

We show the effective cross section $\sigma_{{\rm  eff},\nu_j\bar{\nu}_j}$ for $\nu_j\bar{\nu}_j$ collisions in Fig.~\ref{EffectiveCrossSection} and $m_{\nu_j} = 0.0218\;{\rm eV}$, which is our lightest neutrino mass eigenstate. 
This channel includes $s-$ channel annihilation into fermion pairs (Eq.~\ref{schannelannihilation}), as well as $t-$channel contributions from charged lepton and neutrino pair production. Here, we see the thermal broadening increasing the resonance width at the cost of reducing the cross section at resonance energy. This effect is moderate at $z_s=10$ and is significant at $z_s=100$. As C$\nu$B neutrinos become relativistic, the peak of the effective cross section is shifted to lower energies due to the $1/(1+z)$ scaling of the resonance energy (see Eq.~\ref{Rel_Resonance}). The effect of cross section broadening is less prominent for channels other than the resonant neutrino-antineutrino annihilation, as their cross sections are smoother functions of $s$. However, the shift of the cross section to lower energies at high redshifts is present in all channels. 

\begin{figure}
\includegraphics[width=0.45\textwidth]{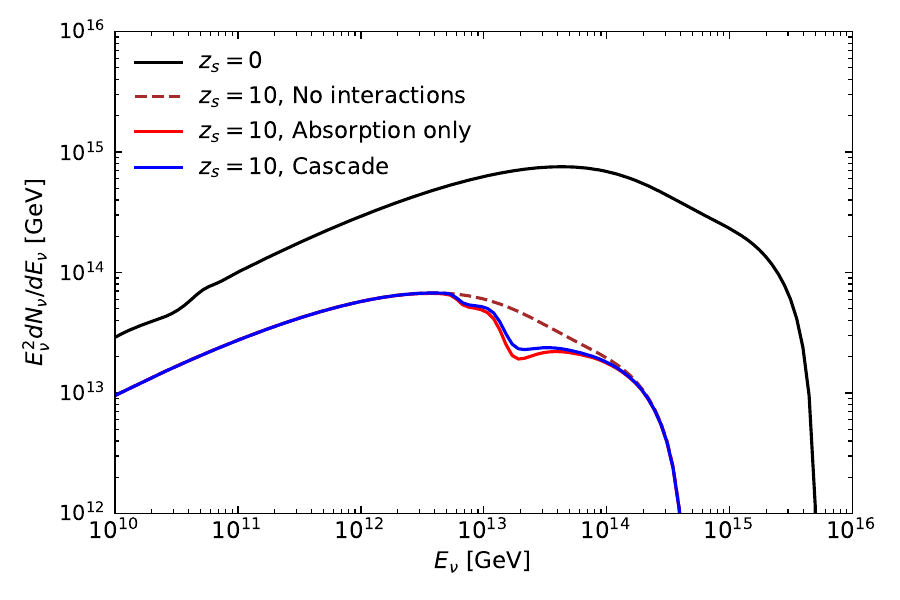}
\caption{Effects of neutrino interactions with the C$\nu$B on propagated neutrino spectra, arising from $\chi\to b\bar{b}$ decays with $m_{\rm DM}=10^{16}$~GeV, which decay at $z_0=10$. The black curve is the spectrum from the decay of a VHDM particle. The dashed brown curve corresponds to the observed spectrum from the decay of a VHDM particle at $z_s=10$, accounting for adiabatic losses during the cosmic expansion. The red (blue) curve includes the effects of neutrino absorption (absorption and cascade). The injected flux at $z_s=0$ is normalized such that the total energy is $m_{\rm DM}$. 
}
\label{ZBoson_SpectrumImpact}
\end{figure}

We show the effects of resonant $Z$-boson production on the propagated neutrino fluxes from redshift $z=10$ in Fig.~\ref{ZBoson_SpectrumImpact}, for a single $m_{\rm DM} = 10^{16}$~GeV particle decaying at $z_s=10$. 
The absorption effect, denoted by the red curve, results in two dips, at $E_\nu\sim 7\times10^{12}~{\rm GeV}$ and $E_\nu\sim 2\times10^{13}~{\rm GeV}$, which corresponds to the different neutrino masses. We would not observe two separate dips for the masses 0.0218 eV and 0.0235 eV, as their cross sections would be smeared out by the thermal distribution of the C$\nu$B (see Figure \ref{EffectiveCrossSection}). 
The neutrino cascade is responsible for the partial reinjection of neutrinos in the $5\times 10^{12}~{\rm GeV}- 7\times 10^{13}~{\rm GeV}$ energy range. Note that there is an overall loss of neutrinos because the $Z$ boson produced during resonance decays to neutrinos $\sim20\%$ of the time. 

\begin{figure*}
    \centering
    \includegraphics[width=0.31\textwidth]{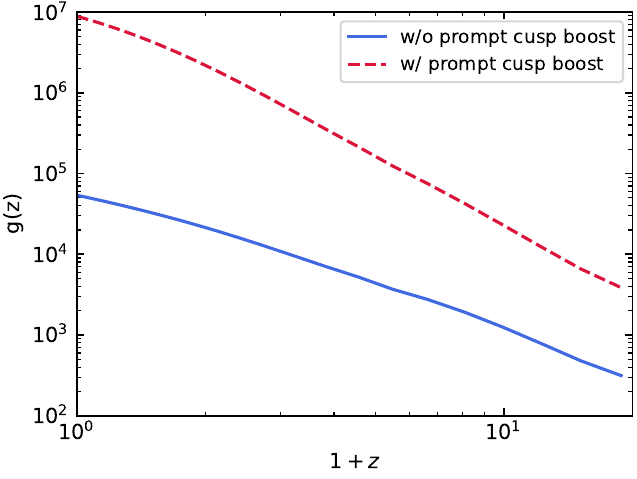}%
    \includegraphics[width=0.33\textwidth]{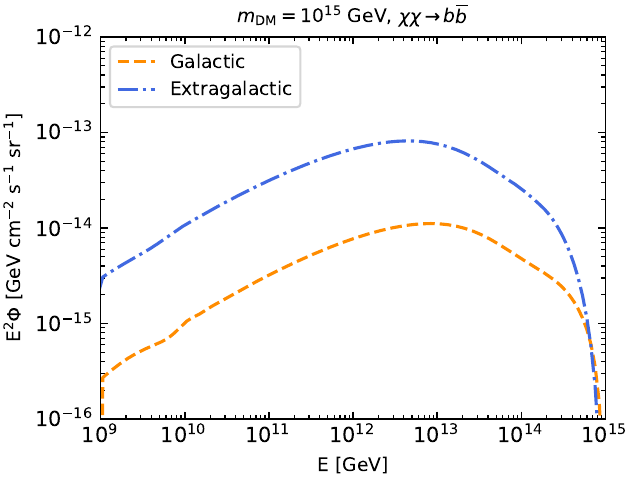}%
    \includegraphics[width=0.33\textwidth]{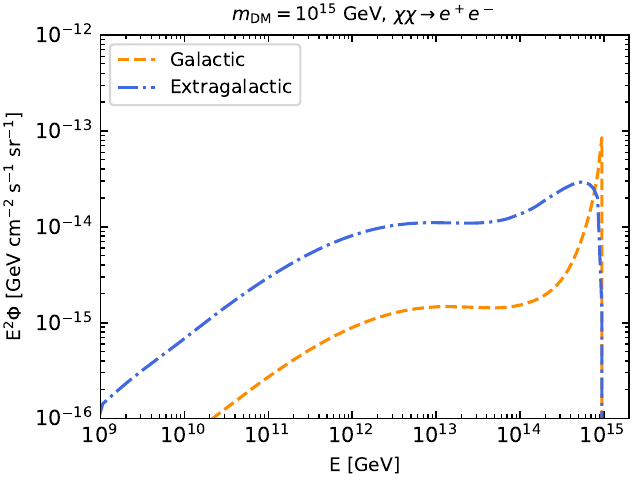}
    \caption{Left: The extragalactic flux multiplier $g(z)$ as a function of redshift, using the parametrizations for the halo mass function from Ref.~\citep{Tinker:2008ff} and $c-M$ relation from Ref.~\citep{Correa:2015dva}. We use a minimum halo mass of $10^6M_\odot$. Galactic and extragalactic components are shown individually for the annihilation of $10^{15}$~GeV dark matter and for $b\overline{b}$ (middle) and $e^+e^-$ (right) channels, corresponding to a fiducial value of cross section $\langle \sigma v \rangle = 10^{-19}$ cm$^3$s$^{-1}$. The enhancement in the extragalactic component arises due to the effect of prompt cusps.}
    \label{fig:gz}
\end{figure*}

\section{\label{sec:nuvhdm}UHE$\nu$s from VHDM}
We consider DM decay (or annihilation) to neutrino pairs $\nu_\alpha+\overline{\nu}_\alpha$ through the channel $X\overline X$, where $X$ is an arbitrary standard model particle. For a VHDM particle of mass $m_{\rm DM}$, each neutrino carries energy $m_{\rm DM}/2$ (or, $m_{\rm DM}$) for decay (or annihilation). 

Let $dS_{\nu_\alpha}/d\varepsilon$ ($dS_{\bar{\nu}_\alpha}/d\varepsilon$) be the prompt $\nu_\alpha(\bar{\nu}_\alpha)$ spectrum from a single DM decay/annihilation process. Considering a single decay (annihilation) occurring at a redshift $z_s$, we can compute the observed spectrum $dN^{\rm dec(ann)}_A/dE (z_s)$, after propagation is taken into account (i.e., at $z=0$), by solving Eq.~\ref{TransportEquation_Time} for $\tilde{N}^{\rm dec(ann)}_A$, assuming $\tilde{\mathcal{Q}}_A(\varepsilon,t) = \delta(t-t(z_s)) dS^{\rm dec(ann)}_A/d\varepsilon (\varepsilon)$. Here $dS^{\rm dec(ann)}_A/d\varepsilon$ is the prompt spectrum in the mass eigenstate basis, obtained by using Eq.~\ref{FlavorToMassConversion}. Note that we use $E$ to denote the observed energy, which is the comoving energy at redshift zero, $E=\varepsilon(z = 0)$. The spectra $dN_A/dE (z_s)$ can be converted back to per-flavor spectra by inverting Eq.~\ref{FlavorToMassConversion}. Details on how to solve Eq.~\ref{TransportEquation_Time} numerically can be found in the Appendix.

The all-flavor differential neutrino and antineutrino flux per unit solid angle and area for decaying or annihilating dark matter can be written as
\begin{align}
\Phi(E) = \Phi_{\rm G}(E) +\Phi_{\rm EG}(E) 
\end{align}
where $\Phi_{\rm G}(E)$ and $\Phi_{\rm EG}(E)$ are the Galactic and extragalactic contributions at Earth. 

We calculate the Galactic contribution to the neutrino flux by integrating over the DM density distribution along the line of sight. The Galactic flux of decaying VHDM at Earth is~\cite{Murase:2012xs} 
\begin{align}
    \Phi^{\rm dec}_{\rm G}(E)= \dfrac{1}{4\pi m_{\rm DM} \tau_{\rm DM}} \dfrac{dS^{\rm dec}}{dE}  \times \int_0^{s_{\rm max}(\theta)}\rho_{\rm DM} (R(s)) ds
\end{align}
where $\theta$ is the angle between the line-of-sight vector and the Galactic center direction, and $dS^{\rm dec}/dE$ is the all-flavor prompt neutrino spectrum from VHDM decay. Neutrino propagation effects are negligible for the Galactic component. The boundary of the halo in the direction $\theta$ is given as
\begin{equation}
s_{\rm max}(\theta) = R_{\rm sc}\cos\theta + \sqrt{R_h^2 - R_{\rm sc}^2\sin^2\theta}.
\end{equation}
We take $R_h=100$ kpc as the size of the Galactic halo \cite{Aloisio:2015lva} and $R_{\rm sc}=8.34$ kpc as the distance between the Sun and the Galactic center.
 
The neutrino flux from the decay of extragalactic VHDM is given as~\cite{Murase:2012xs} 
\begin{align}
    \Phi^{\rm dec}_{\rm EG}(E)=
    \dfrac{\Omega_{\rm DM}\rho_c}{4\pi m_{\rm DM}}\int dz \bigg|\dfrac{dt}{dz}\bigg| F(z) \dfrac{dN^{\rm dec}}{dE}(z_s=z),
\end{align}
where $dN^{\rm dec}/dE$ is the all-flavor neutrino spectrum observed at Earth for the extragalactic component originating from redshift $z$ (e.g., Eq.~2.3 of \cite{Murase:2012df}),  which accounts for attenuation and cascade effects due to extragalactic propagation, $\rho_c$ is the critical density in a flat Friedmann–Lemaître–Robertson–Walker universe and $\rho_{\rm DM} = \Omega_{\rm DM}\rho_c$. We take $\Omega_{\rm DM} h^2=0.113$ and $\rho_c h^{-2}=1.05\times10^{-5}$~GeV~cm$^{-3}$, where $h=0.673$ is the dimensionless Hubble constant \citep{Planck:2018vyg}. 
The factor $F(z)$ is the redshift evolution of the source population for extragalactic dark matter, which is considered unity for the decay case.

For annihilating dark matter (${\rm DM}+{\rm DM}\rightarrow X\overline{X}$), the Galactic flux is proportional to the square of density and the velocity averaged cross section $\langle\sigma v\rangle_{\rm DM}$,
\begin{align}
    \Phi^{\rm ann}_{\rm G} (E, \theta) &= \dfrac{1}{4\pi}\dfrac{\langle\sigma v\rangle_{\rm DM}}{2 m^2_{\rm DM}} \dfrac{dS^{\rm ann}}{dE}\int_0^{s_{\rm max}(\theta)}\rho^2_{\rm DM} (R(s)) ds
\end{align}

The extragalactic flux from VHDM annihilation is enhanced due to the clustering of dark matter in subhalos and is represented as 
\begin{align}\nonumber
\Phi^{\rm ann}_{\rm EG}=& \dfrac{\langle \sigma v\rangle_{\rm DM}(\Omega_{\rm DM}\rho_c)^2}{8\pi m^2_{{\rm DM}}}\\
  &  \times \int dz \bigg|\dfrac{dt}{dz}\bigg| \dfrac{dN^{\rm ann}}{dE}(z_s=z)  g(z)(1+z)^3 
\end{align}
where we calculate the flux multiplier $g(z)$ as
\begin{equation}
g(z) = \int dM \dfrac{dn_{\rm halo}}{dM} \hat{g}(c(M,z)) \dfrac{M}{\rho_{\rm DM}}\dfrac{\Delta_c}{\Omega_{\rm DM}}.
\label{eqn:gz}
\end{equation}
Here $V(dn_{\rm halo}/dM)\delta M$ is the number of halos contained in the mass range $M$ to $M+\delta M$, assuming halos are virialized out to radius $r_v$ corresponding to a cosmological volume V, and $\Delta_c$ is the virial overdensity compared to the critical density, taken to be 200. The quantity $\hat{g}(c(M,z))$ is the flux multiplier of a single halo with concentration $c=c(M,z)$. We use an updated expression for the $c-M$ relation from Ref.~\citep{Correa:2015dva} (cf. Appendix B, therein), provided for the \textit{Planck} cosmology \citep{Planck:2018vyg}.

The functional form of $\hat{g}(c)$ is determined by the DM density profile, which we take to be the Navarro–Frenk–White (NFW) profile~\cite{Navarro:1996gj, Navarro:2003ew}, given as $\rho_{\rm NFW} (R) = {\rho_0}/[{(R/R_s)(1+R/R_s)^{2}}$]. We take the scale radius $R_s=11$~kpc, and the dark matter density in the solar neighborhood to be $\rho_{\rm sc} c^2=0.43$~GeV/cm$^3$ \cite{Karukes:2019jxv}. For a more diffuse DM density distribution, such as the Burkert profile \citep{Burkert:1995yz}, the Galactic component of the annihilation fluxes may be significantly reduced. On the other hand, the concentration parameter depends on the mass accretion history, and only halos whose accretion histories deviate strongly from the NFW shape would have mass profiles with notable differences in the clumping factor \citep{Ludlow:2013bd, Ng:2013xha}. The halo mass function $dn_{\rm halo}/dM$ can be simply estimated using the Press-Schechter formalism \cite{Press_1974} or the extended Press-Schechter formalism \cite{Sheth:1999mn, Sheth:2001dp}, which can also be parameterized with numerical simulations~\cite{Jenkins:2000bv,Reed:2003sq,Tinker:2008ff}. In our calculations, we use the parametrization provided by Ref.~\cite{Tinker:2008ff}, calculated using the open source code HMFcalc~\cite{Murray:2013qza}\footnote{\url{https://github.com/halomod/hmf}}. A minimum mass of $10^{8} h^{-1} M_\odot $ gives $g_0= g(z=0)\simeq 10^{4}$. Extending to masses as low as $10^{-6}h^{-1} M_\odot$ gives $g_0\simeq 2\times10^{5}$. 

There can be an additional boost coming from prompt cusps that are formed by the collapse of peaks in the smooth initial density field during dark matter formation in the early Universe~\cite{Delos:2022bhp, Delos:2023ipo, Ondaro-Mallea:2023qat}. They have strongly peaked dark matter densities, and although their contribution to the total dark matter content is insignificant, their annihilation can substantially influence the flux of multimessenger signals. The annihilation rate per dark matter mass can be expressed as $\rho_{\rm eff}\langle \sigma v \rangle_{\rm DM}/2m_{\rm DM}^2$, where $\rho_{\rm eff}$ is given by Eq.~2.7 in Ref.~\citep{Delos:2022bhp}. For our work, we consider halos with $>10^{6}M_\odot$, and convolute the integrand in Eq.~\ref{eqn:gz} with the boost factor $B(M)$ due to the prompt cusp. The latter has been calculated in Ref.~\citep{Delos:2022bhp} for the cases with and without tidal stripping effects and for weakly interacting massive particles (WIMP) with $m_{\rm DM}=100$~GeV, in which the decoupling from radiation occurs at a temperature of $T_{\rm kd}=30$~MeV. VHDM is cold, and the decoupling may occur at much higher temperatures, so our assumption about $\rho_{\rm eff}$ may still hold.

We show the flux multiplier $g(z)$ in the left panel of Fig.~\ref{fig:gz} with and without the additional boost due to the prompt cusp effect. We calculate the limits on annihilation cross section that incorporates the boost due to prompt cusp formation and thus improve upon earlier estimates in the literature. The middle and right panels in Fig.~\ref{fig:gz} show that the extragalactic component dominates in the case of annihilation for the values of flux multiplier $g(z)$ considered in this study (left panel). We assume a fiducial value of the velocity-averaged cross section $\langle \sigma v \rangle=10^{-19}$ cm$^3$s$^{-1}$ for these figures. The value is chosen within a reasonable range to explain the underlying phenomenon and present the spectral shape. The actual values obtained from our analysis are presented in Sec.~\ref{sec:results}. Thus, our VHDM limits are primarily constrained by the extragalactic component for annihilation, whereas in the case of decay, the Galactic component is dominant. Although there can be contributions from Galactic cusps as well, the extragalactic flux is more crucial than the Galactic flux in constraining the annihilation cross section for large values of $g_0$.

\section{\label{sec:results}Results\protect}

\begin{figure*}
    \centering
    \includegraphics[width=0.4\textwidth]{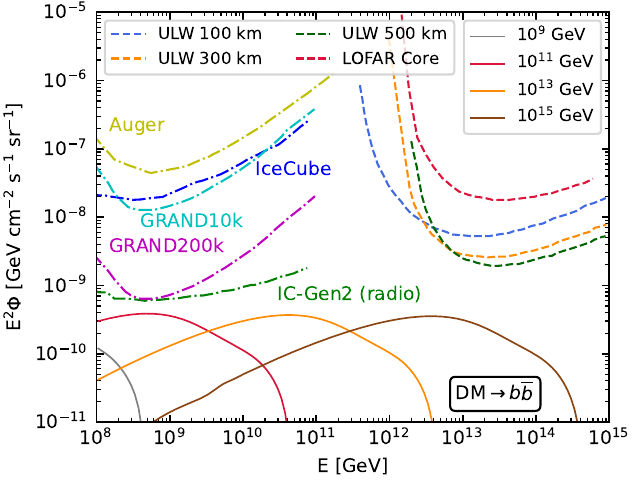}%
    \includegraphics[width=0.4\textwidth]{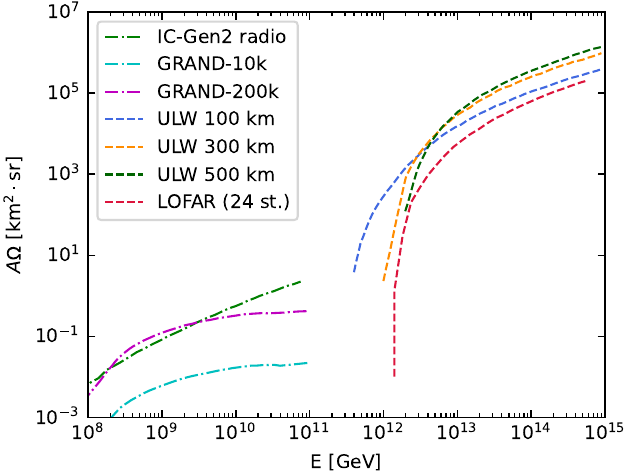}
    \caption{\textit{Left:} Total (Galactic + extragalactic) neutrino flux from the decay of dark matter of different masses through the $b\bar{b}$ channel with $\tau_{\rm DM}=5\times10^{29}$~s. We show the most restrictive upper limit (90\% C.L.) from the currently operating detectors IceCube and Auger, and the projected 90\% C.L. sensitivities of IceCube-Gen2 radio, GRAND-10k, and GRAND-200k experiments. \textit{Right:} The exposure of different detectors used in this study, extrapolated to $10^{16}$~GeV. The exposure for ULW \citep{Chen:2023mau} and LOFAR \citep{Krampah:2023mpj} detectors is obtained from the available data, while that for IceCube-Gen2 radio array and GRAND are calculated using Eq.~\ref{eqn:exposure}.
    }
    \label{fig:sample}
\end{figure*}

\subsection{UHE$\nu$ detectors}
We show a representative case of neutrino fluxes from decaying VHDM into the $b\overline{b}$ channel for different values of $m_{\rm DM}$ in Fig.~\ref{fig:sample}, assuming a fiducial value of the decay timescale $\tau_{\rm DM}=5\times 10^{29}$~s. This value is used for reference purposes, and the actual values obtained in our analysis are presented later in this section. Here, we adopt this fiducial value to present the spectral shape in contrast to various detector sensitivities.
We display the 90\% C.L. upper limits from currently operating detectors, viz., Auger \citep{PierreAuger:2019ens} and IceCube \citep{IceCube:2018fhm}. We also show the sensitivity curves for planned detectors, viz., IceCube-Gen2 radio array for 5 years of operation \citep{IceCube:2019pna, IceCube-Gen2:2020qha} and GRAND \citep{GRAND:2018iaj} for 3 years of operation time. As a conservative approach, we use the upper bound in the sensitivity band of IceCube-Gen2 that considers the estimated uncertainties. Together with an instrumented volume of $\sim8$ times larger than that of IceCube, the next-generation IceCube-Gen2 detector will increase the neutrino collection rate by almost an order of magnitude, and the addition of a radio array will extend the energy range by several orders of magnitude. GRAND aims to detect radio emissions from extensive air showers created by UHE particles. It consists of independent sub-arrays of 10,000 antennas (GRAND-10k), with the entire configuration comprised of 200,000 antennas (GRAND-200k) with individual sub-arrays deployed at separate radio quiet geographical locations.

The 3-yr differential flux upper limits for the Askaryan Radio Array (ARA) and the Antarctic Ross Ice-Shelf Antenna Neutrino Array (ARIANNA) experiments are comparable and a factor of few better than the IceCube \citep{Allison:2011wk, Anker:2020lre}. The 3-yr sensitivity of the Probe of Extreme Multi-Messenger Astrophysics (POEMMA) experiment is a factor of a few higher than ARA and ARIANNA at around $10^9$~GeV but has a comparable sensitivity at around $10^{11}$~GeV \citep{POEMMA:2020ykm}.
At $\sim30$~EeV, the sensitivity of the Payload for ultrahigh Energy Observations (PUEO) \citep[][]{PUEO:2020bnn} is expected to be two orders of magnitude more sensitive than its predecessor Antarctic Impulsive Transient Antenna (ANITA) \citep{ANITA:2019wyx}. The Radio Neutrino Observatory in Greenland (RNO-G) \citep{RNO-G:2020rmc} will enable the detection of a continuation of the IceCube astrophysical neutrino flux at high energies, assuming a power-law with similar spectral indices, as predicted in Ref.~\cite{Fang:2017zjf}. The sensitivity of The Beamforming Elevated Array for COsmic Neutrinos (BEACON) may be comparable to that of GRAND-200k, and thus, the proton-dominated UHECR models can be tested within a few years of observation time \citep{Ackermann:2022rqc}. 

At $\lesssim 10^{12}$~GeV, the tightest bounds come from IceCube-Gen2 and GRAND-200k. At higher energies, the projected sensitivities of the lunar orbital radio antennas constrain the VHDM decay rate and annihilation cross section. The projected sensitivities for the ULW radio antenna, shown in the left panel of Fig.~\ref{fig:sample}, correspond to different altitudes as indicated and one-year observation time. LOFAR is sensitive to neutrinos with $\gtrsim 10^{13}$~GeV, and recent results for one-minute observation using six core LOFAR `half' stations indicate a sensitivity that is weaker than all past projections. Using the same sensitivity, the estimated limits for 5000 hrs of observations, considering all 24 full LOFAR stations, indicate a significant improvement \citep{Krampah:2023mpj}.

\begin{figure*}
\centering
    \includegraphics[width=0.32\textwidth]{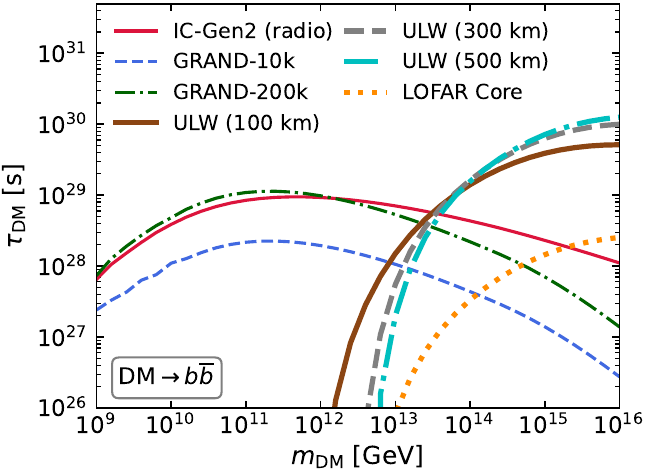}%
    \includegraphics[width=0.32\textwidth]{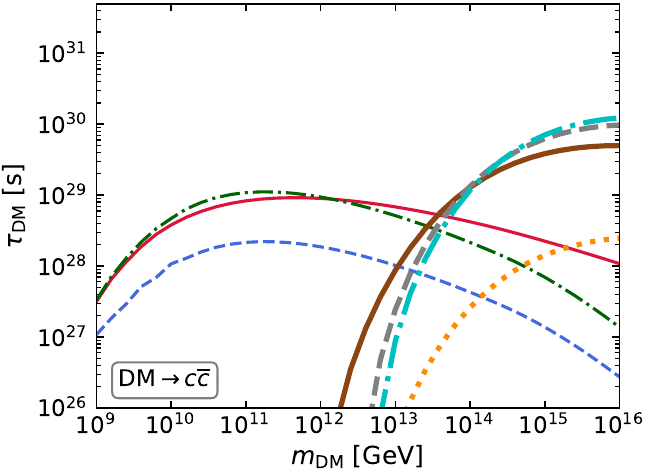}%
    \includegraphics[width=0.32\textwidth]{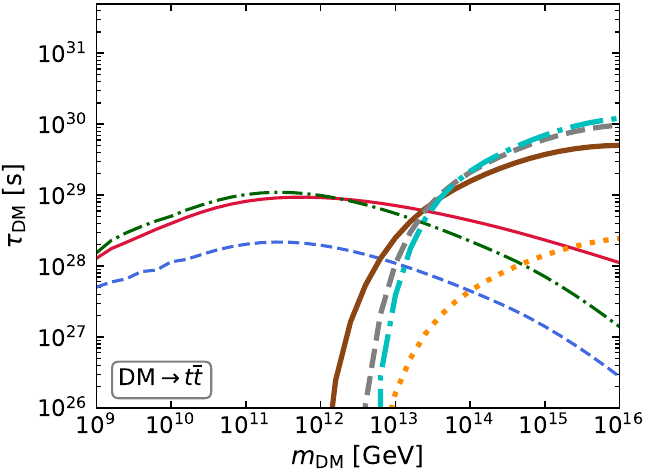}
    \includegraphics[width=0.32\textwidth]{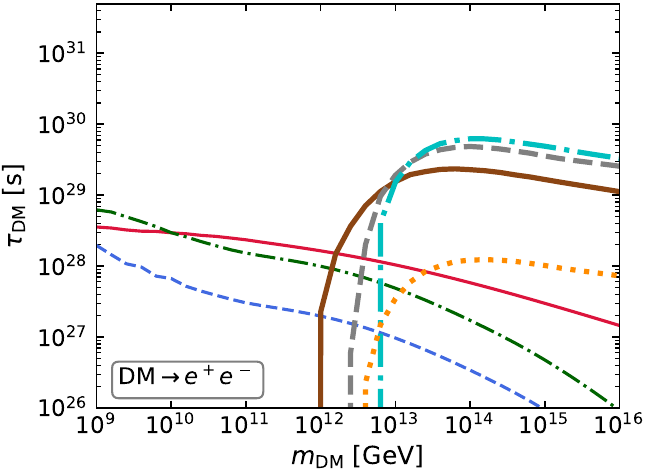}%
    \includegraphics[width=0.32\textwidth]{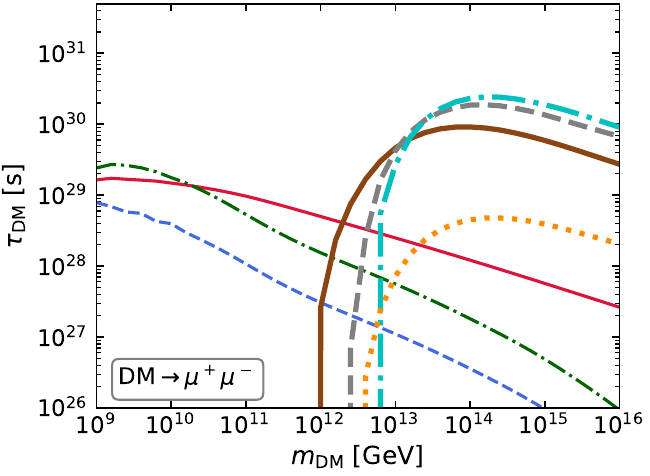}%
    \includegraphics[width=0.32\textwidth]{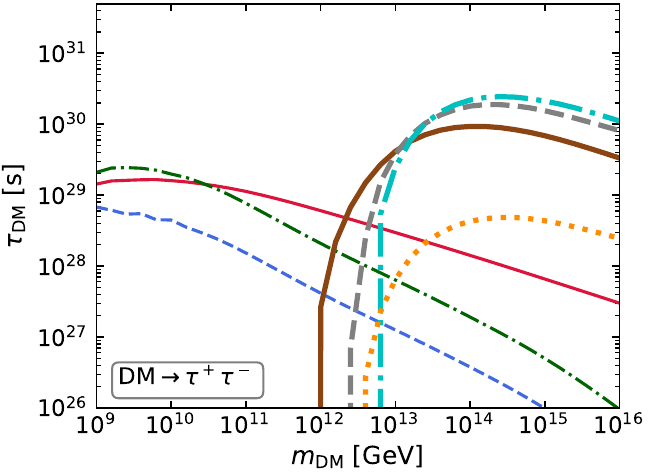}
     \includegraphics[width=0.32\textwidth]{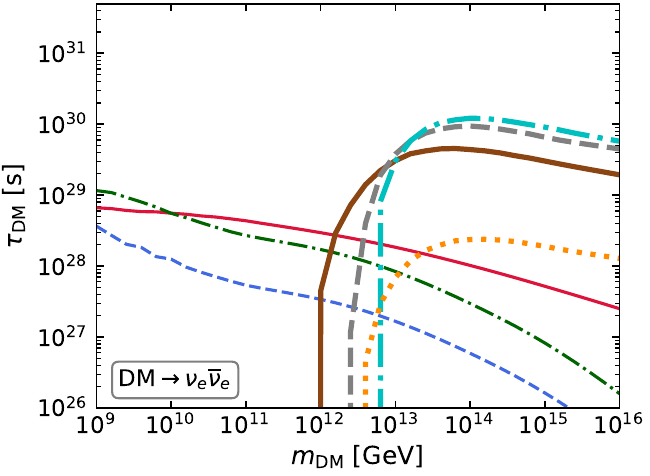}%
     \includegraphics[width=0.32\textwidth]{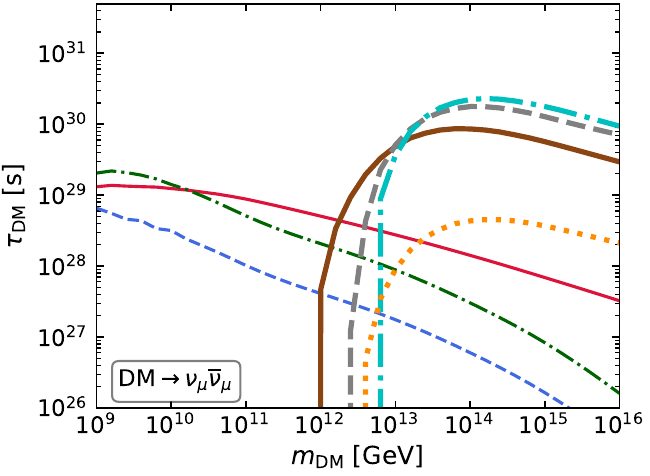}%
     \includegraphics[width=0.32\textwidth]{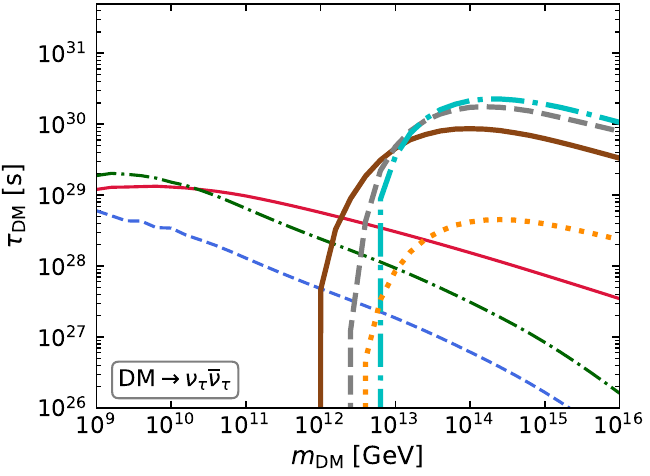}
    \includegraphics[width=0.32\textwidth]{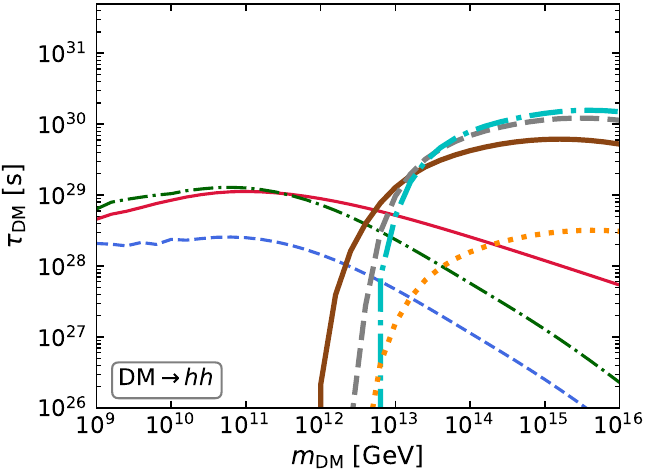}%
    \includegraphics[width=0.32\textwidth]{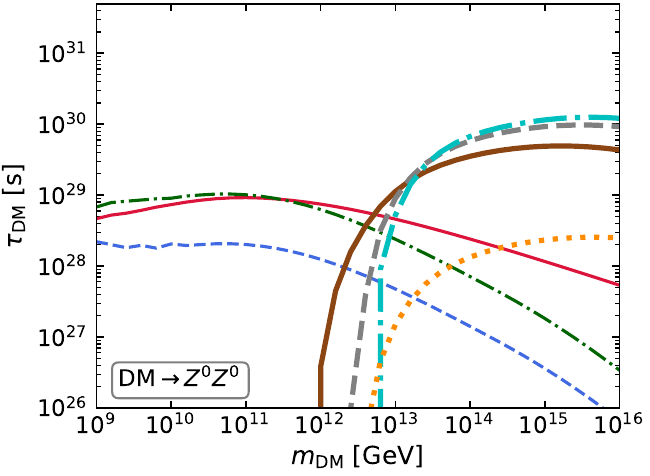}%
    \includegraphics[width=0.32\textwidth]{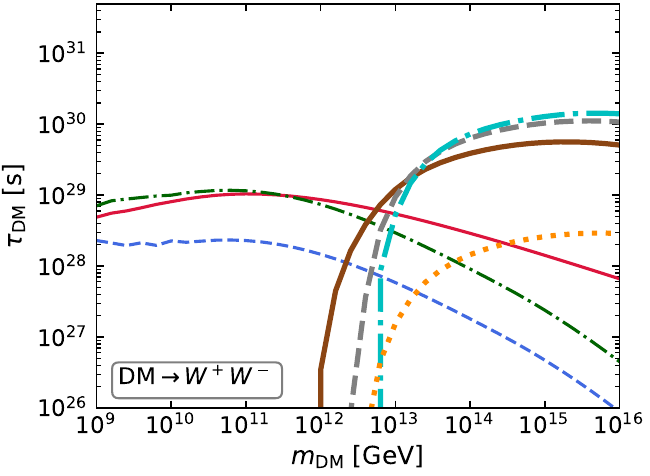}
    \caption{Projected 95\% C.L. limits on the lifetime of VHDM for 5 years of observation time by the future IceCube-Gen2, GRAND, and ULW detectors, and 5000 hrs for LOFAR (see legends). The thick lines correspond to ULW and LOFAR, while the thin lines correspond to IceCube-Gen2 radio and GRAND, respectively.}
    \label{fig:nu_bgd_dec}
\end{figure*}

\begin{figure*}
    \includegraphics[width=0.32\textwidth]{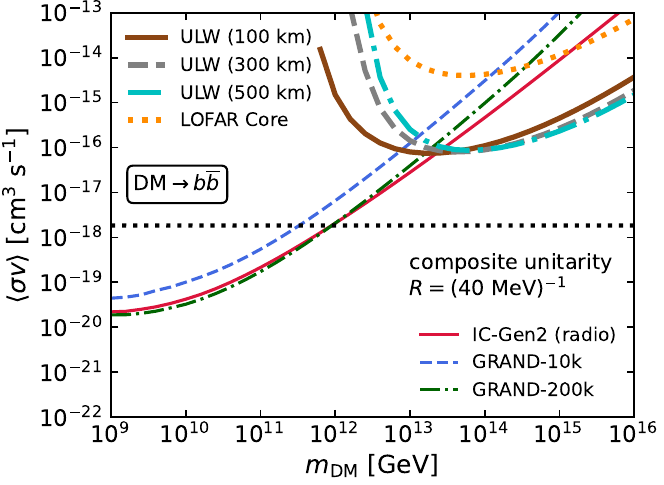}%
    \includegraphics[width=0.32\textwidth]{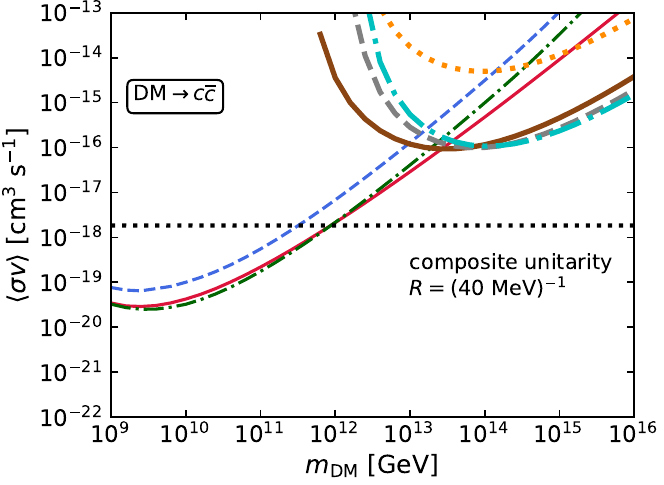}%
    \includegraphics[width=0.32\textwidth]{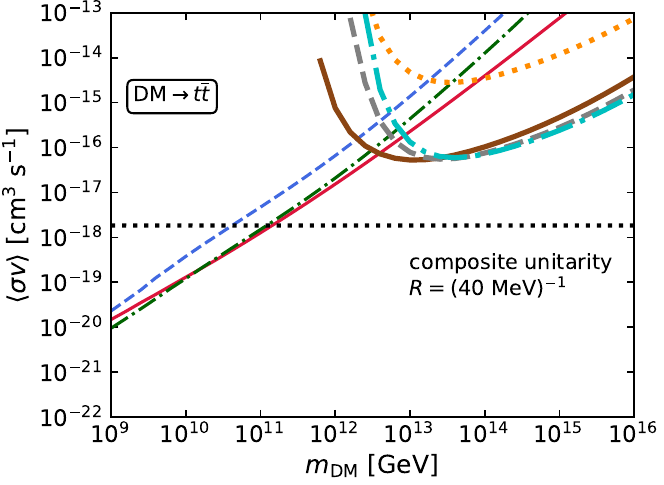}
    \includegraphics[width=0.32\textwidth]{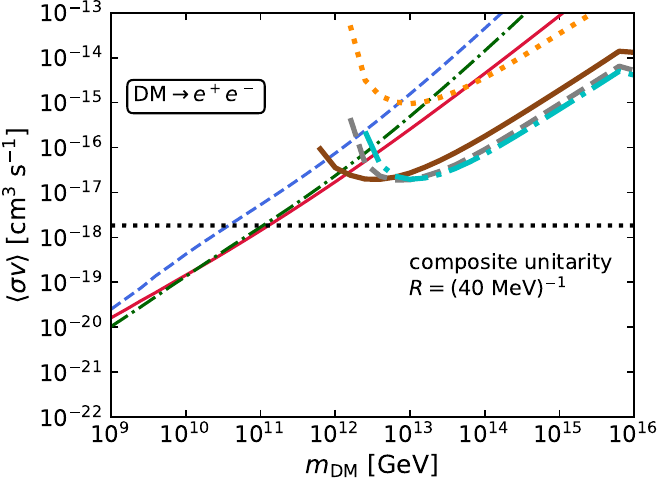}%
    \includegraphics[width=0.32\textwidth]{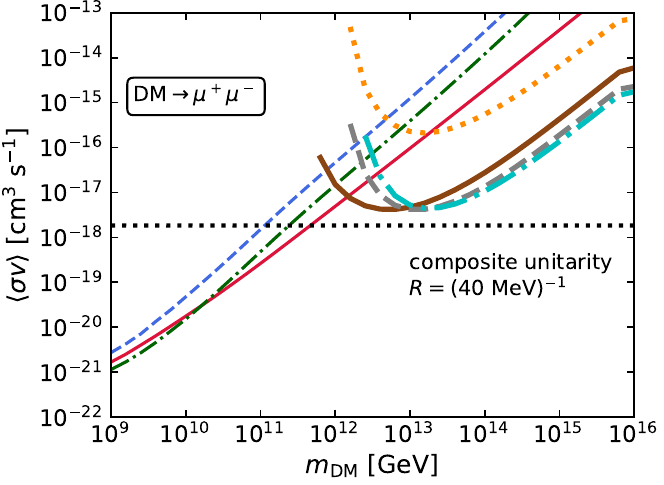}%
    \includegraphics[width=0.32\textwidth]{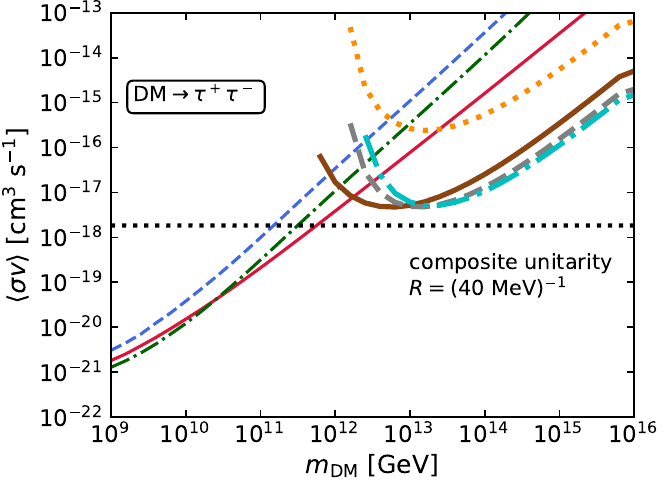}
    \includegraphics[width=0.32\textwidth]{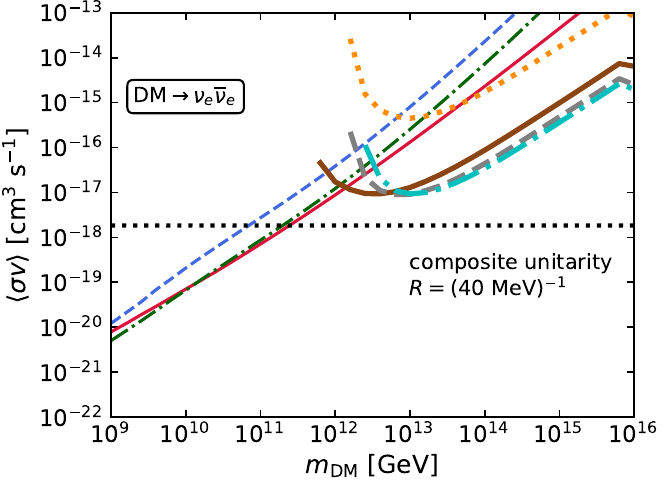}%
    \includegraphics[width=0.32\textwidth]{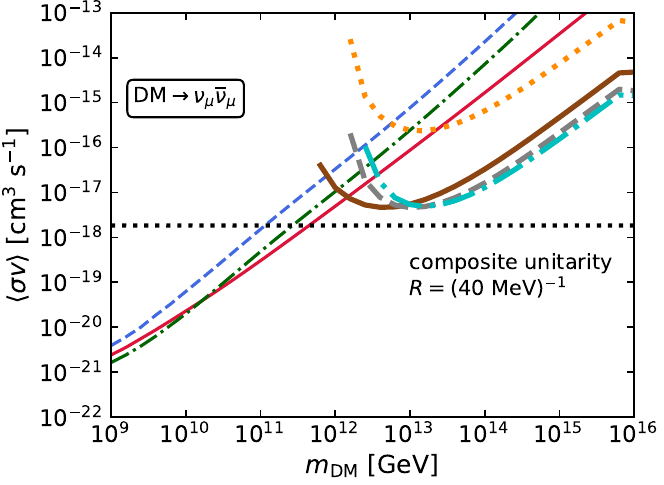}%
    \includegraphics[width=0.32\textwidth]{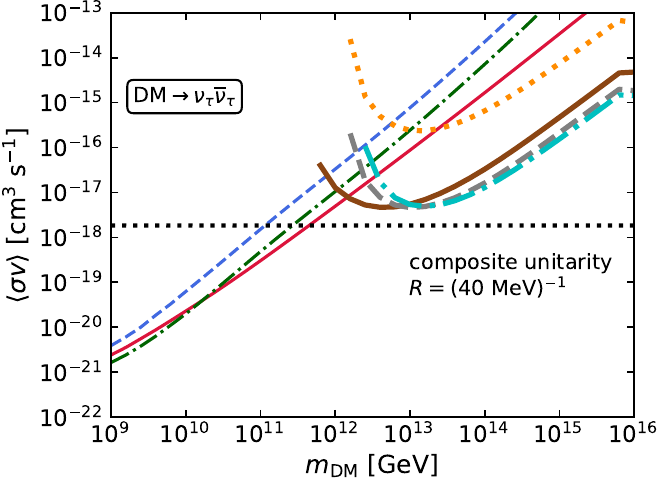}
    \includegraphics[width=0.32\textwidth]{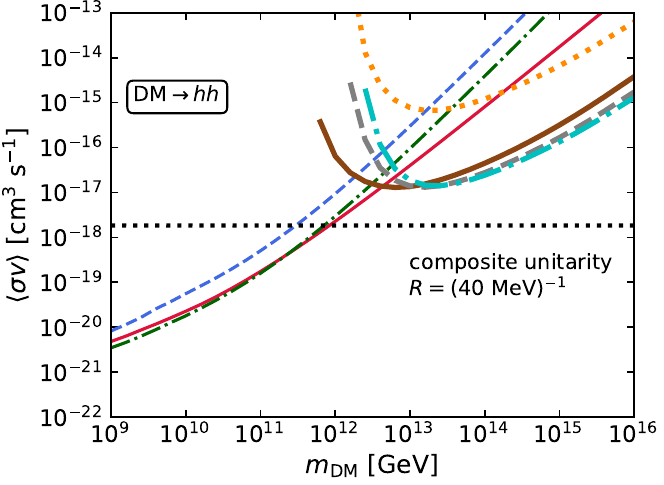}%
    \includegraphics[width=0.32\textwidth]{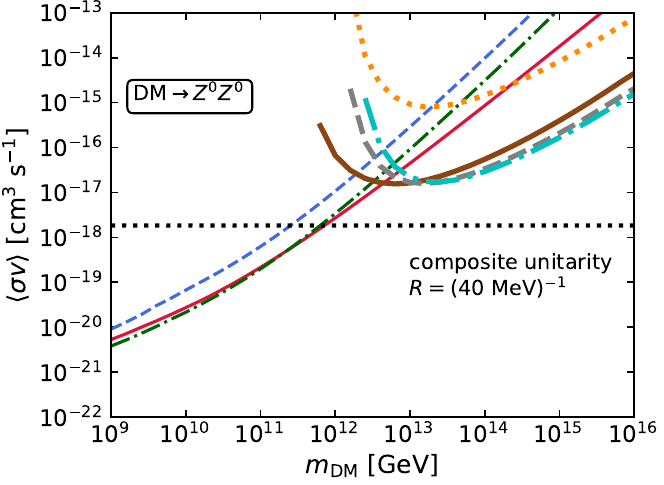}%
    \includegraphics[width=0.32\textwidth]{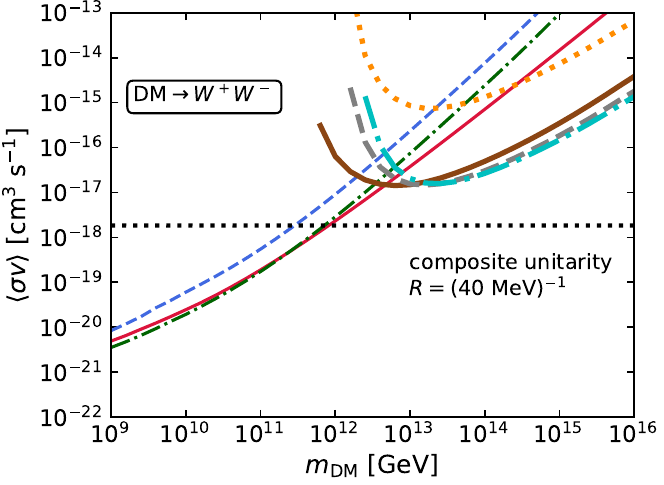}
    \caption{Projected 95\% C.L. limits on the annihilation cross section for 5 years of observation time by IceCube-Gen2, GRAND, and ULW detectors, and for 5000 hrs of observations by LOFAR (see legends).
    The thick lines correspond to ULW and LOFAR, while the thin lines correspond to IceCube-Gen2 radio and GRAND, respectively.}
    \label{fig:nu_bgd_ann}
\end{figure*}

\subsection{Statistical methods}
We calculate the all-flavor UHE$\nu$ fluxes using the formalism described in Sec.~\ref{sec:methods} and Sec.~\ref{sec:nuvhdm} for both decay and annihilation processes and repeat our analysis for various channels. To evaluate the limits on VHDM, we calculate the number of observed neutrino events by convoluting the $\nu$ flux with the effective areas of various neutrino detectors considered in this study, viz., IceCube-Gen2 radio array, GRAND-200k, LOFAR, and the ULW radio telescope. We consider an exposure time of $T_{\rm obs}=5000$~hr for the LOFAR experiment and $T_{\rm obs}=5$~yr for ULW, IceCube-Gen2 radio array and GRAND-200k. The number of neutrino events from dark matter decay or annihilation is given by
\begin{equation}
    \mu_{\rm DM} = T_{\rm obs}\int\int dE d\Omega  \dfrac{d\Phi}{dE} A_{\rm eff}(E, \Omega)
    \label{eqn:fc}
\end{equation}
where $\int d\Omega A_{\rm eff}(E, \Omega)$ is the effective aperture. We use the latest available data for the effective aperture of LOFAR \citep{Krampah:2023mpj} and ULW \citep{Chen:2023mau} detectors. Assuming full mixing due to neutrino oscillations, we estimate the effective area ($A_{\rm eff}$) from the 90\% C.L. flux sensitivity $\Phi_{90}$ of IceCube-Gen2 radio array \citep{IceCube:2019pna, IceCube-Gen2:2020qha} and GRAND-200k \citep{GRAND:2018iaj}, using the expression \citep{IceCube:2010hwb, Reno:2019jtr}
\begin{equation}
A_{\rm eff}(E) = \dfrac{2.44E}{4\pi \ln(10)\Phi_{90}(E)T_{\rm det}},
\label{eqn:exposure}
\end{equation}
corresponding to 2.44 neutrino events for each energy decade, and $T_{\rm det}$ corresponds to the observation time used to calculate the projected sensitivity of the detector. 
The resulting exposures are shown in the right panel of Fig.~\ref{fig:sample}, where we use $T_{\rm det}=5$ years and 3 years for IceCube-Gen2 and GRAND, respectively. We extrapolate the effective areas of these detectors to the highest energies of VHDM particles considered in our study.
We specifically consider three orbital radii for the ULW detector, namely 100~km, 300~km, and 500~km. The sensitivity for detectors at lower orbital radii is lower. 

We adopt the maximum likelihood approach to place constraints on the dark matter decay timescale in the presence of background neutrino flux from astrophysical sources, which is relevant for the most sensitive energy range of IceCube and GRAND experiments. The probability to observe the number of events, $N_{\rm obs}$, follows the Poisson distribution, where the sum of astrophysical and VHDM contributions gives the mean. For an expectation value $\mu_{\rm DM}$ of neutrino events from VHDM, we have 
\begin{equation}
P(N_{\rm obs}\mid \mu_{\rm DM})= \frac{(\mu_{\rm DM}+\mu_{\rm astro})^{N_{\rm obs}}}{N_{\rm obs}!}e^{-(\mu_{\rm DM}+\mu_{\rm astro})},
\end{equation}
where the mean number of events from astrophysical neutrinos is estimated from the background flux model as
\begin{equation}
\mu_{\rm astro}(E_{\rm max}) = T_{\rm obs}\int_0^{E_{\rm max}} dE \dfrac{dN_{\rm astro}}{dE} A_{\rm eff} (E),
\end{equation}
where $dN_{\rm astro}/dE$ is the background neutrino flux of astrophysical origin. Here, we consider neutrinos with energies lower than the maximum energy $E_{\rm max}=m_{\rm DM}/2$ (or, $m_{\rm DM}$) since $E<m_{\rm DM}/2$(or, $m_{\rm DM}$) is the energy range, in which the neutrinos from dark matter decay (or, annihilation) contribute for a given mass. 
Hence, restricting this energy range allows us to test the dark matter signal by minimizing the astrophysical background, resulting in more stringent constraints. We calculate the mean number of astrophysical background events $\mu_{\rm astro}$, collected over a period of $T_{\rm obs} =5$ years, using the projected neutrino sensitivities. For a given dark matter hypothesis, we can also calculate the $\mu_{\rm DM}$, below which the VHDM neutrino signal can be ruled out at $95\%$ C.L. This, in turn, provides the constraints on decay timescale and annihilation cross section from Eq.~\ref{eqn:fc}.

We derive the upper limits on $\mu_{\rm DM}$ following the approach in Refs.~\cite{Cowan:2010js,Chianese:2021htv}. For a given number of detected events $N$, we compute the test statistic (TS),
\begin{equation}
    {\rm TS} = \begin{cases}
        0 &(\mu_{\rm DM}<\hat{\mu}_{\rm DM})\\
        -2\ln\bigg(\dfrac{\mathcal{L}(N|\mu_{\rm DM})}{\mathcal{L}(N|\hat{\mu}_{\rm DM})}\bigg) & (\mu_{\rm DM}\geqslant \hat{\mu}_{\rm DM}),
    \end{cases}
    \label{eqn:likelihood}
\end{equation}
where the likelihood $\mathcal{L}$ is a Poisson distribution and $\hat{\mu}_{\rm DM}$ is the estimator of the mean $\mu_{\rm DM}$ that maximizes $\mathcal{L}$. In the case of Poisson distributions, $\hat{\mu}_{\rm DM} = N-N_{\rm astro}$. 

Each value $\mu_{\rm DM}$ has an associated TS distribution. 
If the experiment observes $N_{\rm obs}$ neutrino events, then we can exclude the DM hypothesis at 95\% C.L. by finding the value of $\mu_{\rm DM}$ for which the $p-$value is equal to 0.05. When setting upper limits, $N_{\rm obs}$ is a random variable with Poisson mean $\mu_{\rm astro}$ (background only), so we can compute the average $\langle \mu_{\rm DM}\rangle$. We use the upper limit $\langle \mu_{\rm DM}\rangle$ to calculate the lower limit on $\tau_{\rm DM}$ (or upper limit on $\langle\sigma v\rangle$) for each value of $m_{\rm DM}$.

Cosmogenic neutrinos can significantly contribute to the neutrino fluxes at energies $\lesssim10^{11}$~GeV. In this work, we use the neutrino flux model derived in Fig. S2 of Ref.~\citep{Fang:2017zjf} to calculate constraints on the decay and annihilation rates, including astrophysical and or cosmogenic backgrounds. This model provides the grand-unification scenario for neutrinos, $\gamma$ rays, and UHECRs~\cite{Murase:2016gly}, and predicts the smooth transition from source neutrinos to cosmogenic neutrinos. It represents the case where the IceCube spectrum is extended to ultrahigh energies, compatible with the nucleus-survival bound for a spectral index of 2.3~\cite{Murase:2010gj}, which may be regarded as a reasonable but optimistic case compared to other cosmogenic neutrino flux predictions \cite{AlvesBatista:2018zui}. Recent observations by Auger indicate a progressively heavier composition with energy. But measurements by Telescope Array observatory prefer light nuclei, indicating a higher proton fraction. However, we are already considering the proton-only case in our choice of the cosmogenic neutrino flux model, which leads to the highest neutrino flux. The model also uses a rather strong redshift evolution $(1+z)^3$ up to $z=1.5$, thus increasing the flux. We chose this optimistic scenario for cosmogenic flux, as it would lead to the most conservative limit on DM decay rates and annihilation.

The neutrino flux in other studies \citep[see, e.g., ][]{vanVliet:2019nse, Heinze:2019jou, AlvesBatista:2018zui} is much lower than that considered for the background in our analysis, leading to stronger constraints. Hence, our estimates of DM decay rate and annihilation timescale are weaker than could have been obtained using the aforementioned models. Indeed, the cosmogenic flux of Ref.~\citep{Fang:2017zjf} is compatible with that obtained in Ref.~\citep{vanVliet:2019nse} and higher than the results obtained from the fitting to the Auger data above the ankle~\cite{Heinze:2019jou}. For the latter, the predicted flux decreases sharply at $\gtrsim 10^{8}$ GeV, with a peak outside our energy range of interest.

In the grand-unification scenario~\citep{Fang:2017zjf}, the authors assume that cosmic rays accelerated in the relativistic jets, from acceleration onto supermassive black holes, can be confined in the Galaxy cluster for 1-10 Gyr, leading to efficient interactions of cosmic-ray nuclei with baryons and infrared background photons in the cluster. The UHECRs that escape the cluster propagate through the extragalactic medium, producing cosmogenic neutrinos. Cumulatively, these two components give rise to a background neutrino flux essential in the energy range of the planned IceCube-Gen2 and GRAND experiments. However, in our energy range of interest at $E_\nu\gtrsim 10^9$ GeV, the $pp$ component is negligibly small, and we effectively use the predicted cosmogenic neutrino flux only for our background model. In addition, the Auger and TA data have been fitted in this model to calculate the cosmogenic neutrino flux while simultaneously conforming to IceCube diffuse astrophysical flux measurements and Fermi-LAT data. The model we use considers a rather strong redshift evolution $(1 + z)^3$ up to $z=1.5$

The background neutrino flux at $\gtrsim 10^{9}$~GeV from astrophysical sources can arise from inside the sources \citep[e.g.,][]{Murase:2009ah,Murase:2014foa,Zhang:2018agl, Carpio:2020wzg} or from interactions of UHECRs. For heavy nuclei-dominated UHECR models, the proton fraction at the highest energies is lower \citep{AlvesBatista:2018zui} and beyond the reach of currently operating detectors. The neutrino flux in most cosmogenic models from UHECR interactions falls off beyond a few times $10^{10}$~GeV. For higher background fluxes, the expected number of neutrino events is higher, and hence, the limits estimated for VHDM decay/annihilation will be weaker. 

While the astrophysical sources are reasonable, the flux may be low, in which case we may get more stringent constraints on VHDM. Assuming no astrophysical neutrino background over the entire energy range of interest, such as that produced inside astrophysical sources or of cosmogenic origin, we calculate projected 95\% C.L. upper limits by requiring the detection of 3.09 events according to the Feldman-Cousins approach \cite{Feldman:1997qc}. The background-free study gives the most stringent limits to the decay timescale and annihilation cross section, as discussed later.

\subsection{\label{subsec:constraints}Constraining VHDM decay and annihilation\protect}

\begin{figure*}
    \centering
    \includegraphics[width=0.32\textwidth]{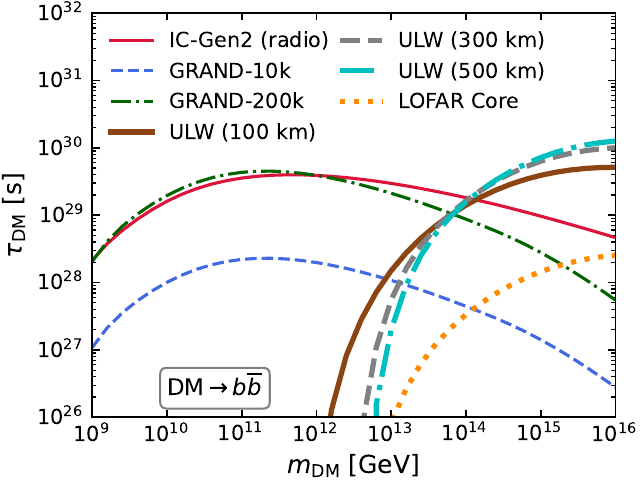}%
    \includegraphics[width=0.32\textwidth]{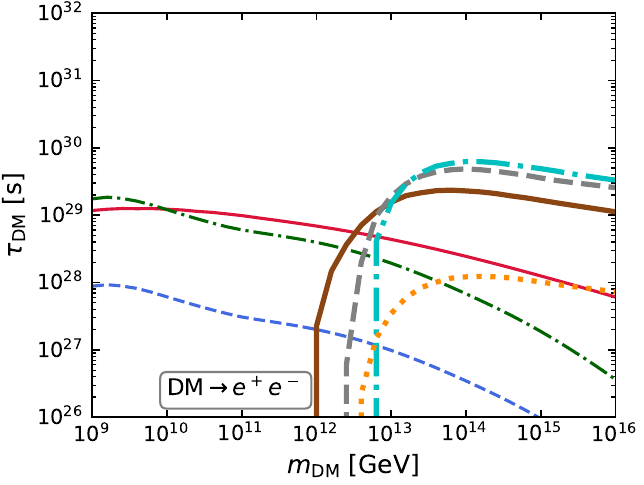}%
    \includegraphics[width=0.32\textwidth]{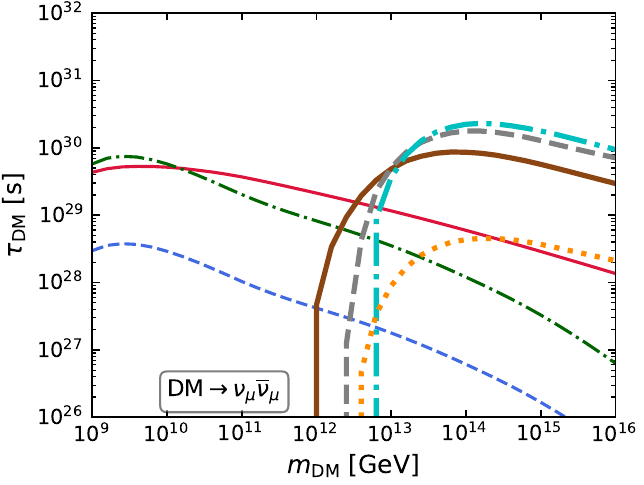}
    \includegraphics[width=0.32\textwidth]{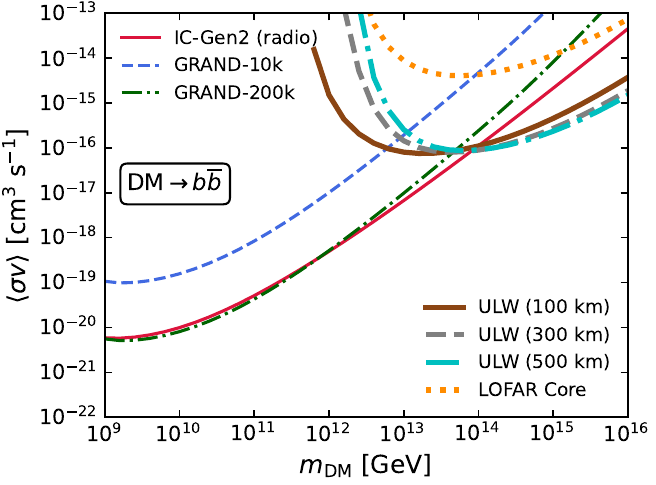}%
    \includegraphics[width=0.32\textwidth]{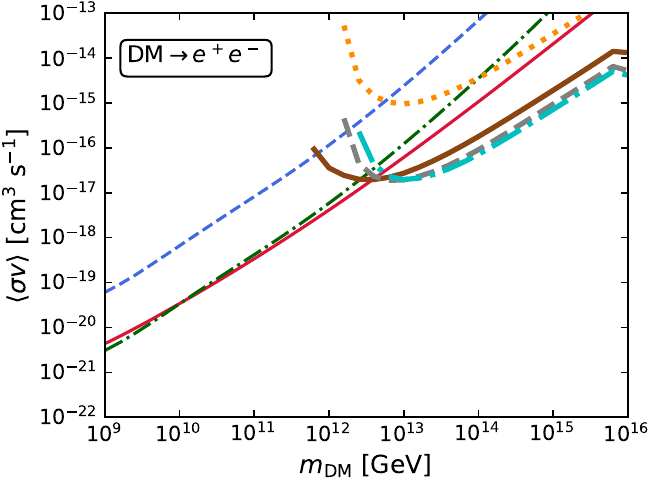}%
    \includegraphics[width=0.32\textwidth]{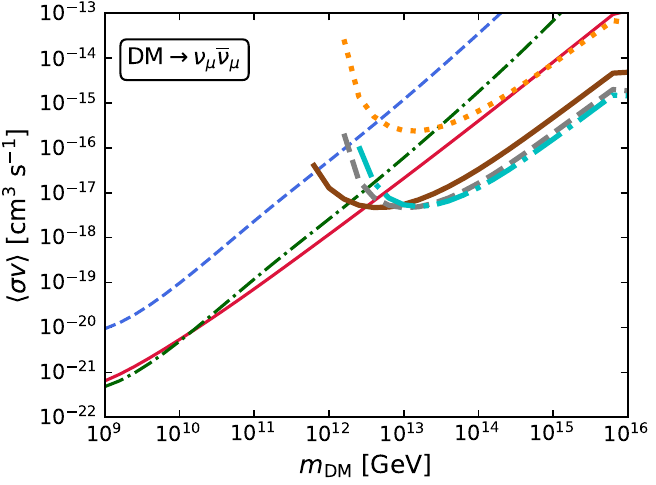}
    \caption{Projected 95\% C.L. upper limits on the VHDM lifetime (upper panel) and annihilation cross section (lower panel) for selected channels, assuming no astrophysical background. We present the limits for 5 years of observation time by IceCube-Gen2, GRAND, and ULW detectors; and for 5000 hrs of observations by LOFAR (see legends).}
    \label{fig:no_bgd}
\end{figure*}
Constraints on VHDM decay and annihilation for various channels are shown in Figs.~\ref{fig:nu_bgd_dec} and \ref{fig:nu_bgd_ann}, respectively, assuming the background contamination by cosmogenic neutrino events according to Ref.~\citep{Fang:2017zjf}. 

For the VHDM decay case, the constraints from IceCube-Gen2 radio array and GRAND-200k provide the best limits at $m_{\rm DM}\lesssim 10^{13}$~GeV for quark and gauge boson decay modes. For the lepton decay modes, the constraints from the ULW detector for an orbital radius of 100~km start becoming more strict than the IceCube-Gen2 radio array and GRAND for VHDM masses of a few times $10^{12}$~GeV. The most stringent limits at the highest energies are estimated for an orbital radius of 500 km and for muon and tau decay channels, constraining the decay timescale to $\tau_{\rm DM}\gtrsim 2\times 10^{30}$~s for $2\times10^{13}$~GeV $\lesssim m_{\rm DM} \lesssim 4\times 10^{15}$~GeV. The limits estimated from muon and tau neutrino channels are also comparable to those estimated from the corresponding charged lepton channels. However, at $m_{\rm DM}\gtrsim 2\times10^{15}$ GeV, we find that the ULW detector can be the most sensitive to neutrino events coming from scalar and vector boson channels, putting a lower bound of $\tau_{\rm DM} \gtrsim 10^{30}$~s. The limits predicted using LOFAR can be significantly higher than IceCube-Gen2 radio and GRAND-200k for the lepton, neutrino, and boson channels for energies beyond a few times $10^{13}$~GeV.

For annihilating dark matter particles going to SM states, the cross section scales as $\propto m_{\rm DM}$. The unitarity bound \citep{Griest_1990} leads to a limit on the cross section or the dark matter mass for a given cross section. The mass is limited to $\lesssim100$~TeV for thermal relics, and in more general it scales as $m_{\rm DM}^{-2}$ (e.g.,~\cite{Murase:2012xs}). However, it can be relaxed if dark matter is not point-like~\citep{Harigaya:2016nlg, Tak:2022vkb}. The dark matter may be composite, and such scenarios have been considered earlier, where dark matter with a large radius arises from strongly coupled confinement theory models in the dark sector \citep{Harigaya:2016nlg}. Although composite particles can evade the $\sim m_{\rm DM}^{-2}$ dependence, composite unitarity may still limit their annihilation rate as \citep{Tak:2022vkb}
\begin{align}
\langle \sigma v \rangle & \leqslant \dfrac{4\pi(1+m_{\rm DM}v_{\rm rel}R)}{m_{\rm DM}^2 v_{\rm rel}} 
\end{align}
where we put $v_{\rm rel}=v = 2\times 10^{-5}$ as an approximate value for the average velocity between VHDM particles in nearby dwarf galaxies \citep{Martinez_2011, McGaugh_2021}. For reference purposes, we show the composite unitarity bound for the minimum value of $R$ possible, such that it is weaker than the limits predicted at $\gtrsim 10^{12}$~GeV for all channels.

The velocity-averaged cross section values of $\langle \sigma v \rangle \gtrsim 10^{-20}$~cm$^3$~s$^{-1}$ can be excluded for $m_{\rm DM}\approx10^{9}$~GeV and for charged lepton, neutrino, and gauge boson annihilation channels. The constraints from LOFAR are more strict than those predicted from IceCube-Gen2 radio and GRAND-200k at $m_{\rm DM}\gtrsim 10^{13}$~GeV for the charged lepton, neutrino, and gauge boson annihilation modes. However, the expected constraints from the ULW mission are significantly more stringent than IceCube-Gen2 radio and GRAND for $m_{\rm DM}\gtrsim 10^{12}$ GeV, limiting $\langle \sigma v\rangle \lesssim 10^{-17}$ cm$^{3}$s$^{-1}$ for $10^{12}$~GeV $\lesssim m_{\rm DM} \lesssim 10^{14}$~GeV for the $\tau^+\tau^-$, $\mu^+\mu^-$ and the corresponding neutrino modes. At the highest mass range in this study, i.e., $m_{\rm DM}\approx10^{16}$~GeV, the vector and scalar boson modes provide the most strict limit of $\langle \sigma v \rangle \lesssim 10^{-15}$~cm$^3$ s$^{-1}$. 

We find that channels resulting in $l\bar{l}$ have neutrino energy spectra with strong peaks at $E_\nu\approx m_{\rm DM}/2$ and $E_\nu\approx m_{\rm DM}$, for decay and annihilation, respectively. When calculating the Galactic components $\Phi_G$ of the VHDM neutrino fluxes, these peaks are present, whereas those in the extragalactic component are smoothened when integrated over redshift. When these peaks coincide with the maximum detector sensitivity energy range, the associated constraints for $l\bar{l}$ channels are more stringent than $q\bar{q}$. This occurs at $m_{\rm DM}\sim 10^{9}$~GeV ($m_{\rm DM}\sim 10^{13}$~GeV) for IceCube-Gen2 and GRAND (ULW and LOFAR). Outside these mass ranges, the sensitivity on $l\bar{l}$ channels drops, and $q\bar{q}$ provides stronger constraints. Such an effect is crucial in the case of constraining VHDM decay timescale, where the Galactic component is dominant. Although the extragalactic component is dominant in the case of annihilation, the peak at the highest energy is still pivotal in placing constraints on the cross section for leptonic and neutrino channels, as shown in the right panel of Fig.~\ref{fig:gz}.

Fig.~\ref{fig:no_bgd} shows the projected limits on the decay lifetime and annihilation cross section without background astrophysical neutrinos for some representative channels. Assuming any detected neutrino signal originates from VHDM, the limits estimated in this case are stricter by a factor of $\sim 4$ than those presented earlier. This is expected since fewer events are expected at the detectors for VHDM-only neutrinos. In particular, the resulting constraints predicted from IceCube and GRAND exposure are more aggressive since the sensitivities peak at around $10^{9}$~GeV and become progressively weaker at higher energies. The energy range for the lunar ULW mission is background-free and thus provides the best limits in this mass range. The limits obtained in this work for GRAND-200k are comparable to the 90\% C.L. limits obtained in \cite{Guepin:2021ljb}, assuming no background. The latter studies the uncertainties related to the dark matter spatial distribution in the Galactic halo. Our limits for decay and annihilation, including the astrophysical background, are therefore weaker. The constraints on annihilation cross section presented in Ref.~\cite{Arguelles:2019ouk} up to $10^{11}$~GeV for the GRAND-200k detector is comparable to that obtained in our work, within a factor of less than two.

\section{\label{sec:discussions}Summary and Discussions\protect}
Observing UHE$\nu$s beyond $10^9$~GeV will provide clues to the origin of the highest energy cosmic rays and also probe the lifetime and annihilation cross section of superheavy dark matter. 
The Auger data, including the UHECR composition, suggest that the background contamination from astrophysical neutrinos would be negligible at energies higher than $\sim10^{11}$~GeV in almost all cosmogenic neutrino models. The detection potential of neutrinos at these energies, employing neutrino interactions on the lunar regolith by, e.g., LOFAR and the proposed lunar ULW radio telescope, can provide powerful constraints at energies beyond ZeV. 

We considered neutrino cascades due to SM $\nu-\nu$ interactions of high-energy neutrinos with the C$\nu$B. The latter effect is more pronounced for neutrinos propagating from redshifts $\gtrsim 5$ and beyond the resonant photomeson production energy ($\approx5\times10^{10}$~GeV at $z=0$) of UHE protons with CMB photons. This is crucial for UHE$\nu$s originating from very high redshifts \citep{PhysRevD.47.5247,Lunardini:2013iwa}.
In the dark matter cases, especially for decaying VHDM, the Galactic component of the neutrino flux dominates over the overall flux obtained at Earth, so this effect is not large. However, the extragalactic flux still undergoes an energy redistribution during their propagation over cosmological distances, which is important if the extragalactic component is dominant by, e.g., early-time decay of VHDM \cite{Ema:2014ufa}. We found that the most stringent lower limits to dark matter decay lifetime can be estimated for lepton channels, indicating a lower limit of a few times $10^{30}$~s at $m_{\rm DM}\gtrsim 10^{13}$~GeV. On the other hand, for annihilating dark matter, the constraints on the velocity averaged cross section can be a few times $10^{-17}$~cm$^3$~s$^{-1}$ at $10^{12}$~GeV $\lesssim m_{\rm DM}\lesssim 10^{14}$~GeV for charged lepton and neutrino channels.

The radio emission from hadronic cascades induced by UHECRs and UHE$\nu$ in the lunar regolith covers a broad range of frequencies and peaks at the GHz regime. The peak intensity of the radiation decreases with decreasing frequencies, but the increased angular spread of the radio emission from the lunar surface makes it efficient for detection by the lunar radio telescope within a broader solid angle.
We found that the constraints from the lunar ULW experiment become more important for dark matter mass $m_{\rm DM}\gtrsim 10^{13}$~GeV. 
In the presence of the astrophysical neutrino background, the limits on $\tau_{\rm DM}$ and $\langle \sigma v\rangle_{\rm DM}$ are weaker, and the detection potential of lunar ULW telescope becomes more important at higher energies. We use the conservative extragalactic neutrino flux model obtained in Ref.~\citep{Fang:2017zjf} assuming the confinement of cosmic ray nuclei in galaxy clusters and galaxy groups, which lead to interactions with the background photons in the cluster. Several other models exist for the astrophysical and cosmogenic neutrino background at EeV energies \citep[see, for e.g.,][]{Engel:2001hd, Ave:2004uj, Seckel:2005cm, Hooper:2004jc, Stanev_2006, Takami:2007pp, Kotera_2010, Moller:2018isk,AlvesBatista:2018zui,Muzio:2023skc}.
The escaping UHECRs from astrophysical sources can interact with the extragalactic background photons to produce cosmogenic neutrinos. The flux of these neutrinos falls off sharply beyond a few times $10^{10}$~GeV, and hence, the constraints put by the lunar radio observations are free from backgrounds and are genuine probes of cosmological neutrino from dark matter decay/annihilation. 

The constraints predicted from neutrinos are complementary to those from UHECR experiments such as Auger, which limits the decay timescale to $\gtrsim 10^{30}$~s for $10^{11}$~GeV $\lesssim m_{\rm DM} \lesssim 10^{15}$ GeV for all decay channels (e.g.,~Ref.~\citep{Das:2023wtk}). The constraints obtained from $\gamma$ rays for boson decay channels in Ref.~\cite{Das:2023wtk} are also weaker at $\gtrsim10^{14}$ GeV than that obtained here using neutrinos. Also, for lepton decay channels, the limits obtained using $\gamma$ rays are weaker than those obtained using neutrinos at energies beyond ${10}^{13}$~GeV. Synchrotron radiation due to $e^\pm$ from VHDM decay in the Galactic magnetic field may also give the lifetime as $\tau_{\rm DM}\gtrsim 10^{31}$~s for $m_{\rm DM}\gtrsim 10^{12}$~ GeV \citep{Munbodh:2024ast} (see also Ref.~\cite{Murase:2012xs} for synchrotron cascades). The Fermi-LAT constraints from observing the Galactic halo indicate decay lifetimes $\gtrsim 10^{27}$~s for the $b\overline{b}$ decay mode and NFW profile \citep{Ackermann_2012}, which is further improved by incorporating electromagnetic cascades \citep{Murase:2012xs,Murase:2012rd,Cohen:2016uyg,Blanco:2018esa, Song:2023xdk,Song:2024vdc}. 
The constraints estimated from cosmic rays give $\tau_{\rm DM}\gtrsim 10^{29}$~s for $10^{11}$~GeV $\lesssim m_{\rm DM}\lesssim 10^{15}$~GeV \citep{Ishiwata:2019aet}.
Imprints of dark matter decay (annihilation) on the CMB provide constraints up to $\sim 10$~TeV, where the limit is $\tau_{\rm DM}\gtrsim 10^{24}$~s ($\langle\sigma v\rangle\lesssim 10^{-23}$~cm$^3$~s$^{-1}$) \cite{Slatyer:2016qyl,Kawasaki:2021etm}. 
For annihilating dark matter, the limit on $\langle\sigma v\rangle$ is proportional to $m_{\rm DM}$. An extrapolation of this limit up to our range of $m_{\rm DM}>10^9$~GeV leads to $\langle\sigma v\rangle\lesssim 10^{16}$~cm$^3$~s$^{-1}$, which is weaker than our result. 

For $m_{\rm DM}\lesssim10^{13}$ GeV, the ground neutrino detectors, viz., IceCube-Gen2 radio array and GRAND-200k, will provide improved constraints compared to the current generation neutrino detectors and hence will play a pivotal role in identifying indirect signatures of VHDM. In addition, other neutrino detectors, KM3NeT \citep{KM3Net:2016zxf}, Baikal-GVD \citep{Baikal-GVD:2019kwy}, and P-ONE \citep{P-ONE:2020ljt} will improve our sensitivity to TeV-PeV neutrinos coming from the southern sky. Proposed detectors such as TAMBO \citep{Romero-Wolf:2020pzh} and Trinity \citep{trinity18} will bridge the gap between the PeV and EeV scale neutrino experiments. Several other ongoing or next-generation neutrino missions, such as BEACON \cite{Southall:2022yil}, POEMMA \citep{POEMMA:2020ykm}, PUEO \citep{PUEO:2020bnn}, RNO-G \citep{RNO-G:2020rmc}, ARIANNA \citep{Anker:2020lre}, ARA \citep{Allison:2011wk}, JEM-EUSO \citep{JEM-EUSO:2007qqf}, TAROGE \citep{Nam:2020hng}, and the AugerPrime upgrade \citep{PierreAuger:2016qzd} of the Pierre Auger Observatory, aim to detect UHE$\nu$s, employing a variety of detection methods such as in-water, in-ice, in-air Cherenkov radiation, as well as fluorescence and particle shower observations. These will provide us with promising probes of VHDM. 

%


\begin{acknowledgments}
We thank Bhupal Dev, Writasree Maitra, Mainak Mukhopadhyay, Deheng Song, and Shigeru Yoshida for the useful discussions. We also thank the GRAND Collaboration, especially Mauricio Bustamante, for reviewing the manuscript and providing useful comments. This research by S.D. and K.M. is supported by KAKENHI No.~20H05852. The work of K.M. is supported by the NSF Grants No.~AST-2108466, No.~AST-2108467, and No.~2308021. J.C. also acknowledges the Nevada Center for Astrophysics and NASA award 80NSSC23M0104 for support.
\end{acknowledgments}

\appendix

\section{\label{appendix:implicitmethod} Details of neutrino transport in space}
The transport equation given by Eq.~\ref{TransportEquation_Time} is solved by treating the redshift energy loss and particle interactions separately. Ignoring redshift energy losses, we first discretize the transport equation as
\begin{eqnarray}\nonumber
    \frac{\partial}{\partial t}\tilde{N}_{A,i} &=& -\tilde{N}_{A,i}\mathcal{A}_{A,i}
    +  \sum_{j\geq i}\tilde{N}_{A,j}\mathcal{B}_{A,j\to i}\Delta\varepsilon_j\\
    & +&\sum_{B\neq A}\sum_{j\geq i}\tilde{N}_{B,j}\mathcal{C}_{B,j\to A,i}\Delta\varepsilon_j+\tilde{\mathcal{Q}}_{A,i},
\end{eqnarray}
where indices $i$ and $j$ correspond to different energy bins and $\Delta\varepsilon_j$ is the bin width. Let $\tilde{N}_{A,i}^{m}$ be the value of $\tilde{N}_{A,i}$ at time $t_m$. Then, the first-order implicit scheme yields
\begin{eqnarray}\nonumber
\frac{\tilde{N}_{A,i}^{m+1}-\tilde{N}_{A,i}^m}{\Delta t}& =& -\tilde{N}_{A,i}^{m+1}\mathcal{A}_{A,i}
    +  \sum_{j\geq i}\tilde{N}_{A,j}^{m+1}\mathcal{B}_{A,j\to i}\Delta\varepsilon_j\\\nonumber
    & +&\sum_{B\neq A}\sum_{j\geq i}\tilde{N}_{B,j}^{m+1}\mathcal{C}_{B,j\to A,i}\Delta\varepsilon_j\\
    &+&\tilde{\mathcal{Q}}_{A,i}^m,
    \label{DiscretizedTransport}
\end{eqnarray}
where $t_{m+1} = t_m + \Delta t$, for some chosen time step $\Delta t$. The same implicit method has been used in the Astrophysical Multimessenger Emission Simulator (AMES)~\cite{Murase:2009ah,Murase:2011yw,Murase:2017pfe,Zhang:2023ewt} (see also Ref.~\cite{Lee:1996fp}). 
The cascade step is solved by finding the values $\tilde{N}_{A,i}^{m+1}$ satisfying Eq.~\ref{DiscretizedTransport}.
To account for redshift energy loss, we choose step sizes $\Delta z = z_{m+1}(t_{m+1})-z_m(t_m)$ such that $1-\Delta z/(1+z_m) = \varepsilon_{i+1}/\varepsilon_i$ \cite{Lee:1996fp} The right hand side is a constant for logarithmically-spaced bins. With this $\Delta z$, all particles at the energy bin $E_{i+1}$ will redshift to the bin $E_i$, and the corresponding expression is
\begin{equation}
\left.\tilde{N}_{A,i}^{m+1}\right|_{\rm with\; redshift\; loss} = \tilde{N}_{A,{i+1}}^{m+1}.
\label{RedshiftSubstitution}
\end{equation}
The step size $\Delta z$ may amount to a long distance step (i.e. comparable to the mean free path). 
Thus, a full propagation step proceeds as follows: We calculate $\Delta z$ and then divide $\Delta z$ into finer intervals, performing several cascade steps on these smaller intervals via Eq.~\ref{DiscretizedTransport}. Finally, we apply redshift energy loss using Eq.~\ref{RedshiftSubstitution}.

Notice that it may not be possible to propagate all the way to $z=0$ using these specially chosen redshift steps. In that situation, the final redshift step may be done via

\begin{equation}
\left.\tilde{N}_{A,i}^{m+1}\right|_{\rm with\; redshift\; loss} = \frac{1+z_m}{1+z_{m+1}}\tilde{N}_{A}^{m+1}\left(\frac{1+z_m}{1+z_{m+1}}\varepsilon_i\right).
\end{equation}

Here, the value of $\tilde{N}_{A}^{m+1}((1+z_m)\varepsilon_i/(1+z_{m+1}))$ is obtained via interpolation.

Next, we outline calculations of the interaction terms in Eq.~\ref{TransportEquation_Time} and their discretized version in Eq.~\ref{DiscretizedTransport}. We go through the terms in order. The total interaction rate is
\begin{equation}
\mathcal{A}_{A}(\varepsilon) = c\sum_{B}\int d^3\mathbf{p}_{B}\frac{dn_B^{{\rm C}\nu{\rm B}}}{
d^3\mathbf{p}_B}(1-\beta_B\cos\theta_{B})\sigma_{AB}(s),
\label{AbsorptionTermA}
\end{equation}
where $\beta_B$ is the target particle's speed, $\theta_{B}$ is the angle between the momenta of particle $A$ and $B$, and $\sigma_{A B}$ is the total cross section for $A$+$B$ collisions, as a function of $s=2p_A\cdot p_B$.

We calculate $\mathcal{B}$ via the expression
\begin{eqnarray}\nonumber
    \mathcal{B}_{A\to A}(\varepsilon) = c \sum_{B,C} & & \int  d^3\mathbf{p}_B \frac{dn_B^{{\rm C}\nu{\rm B}}}{d^3\mathbf{p}_B} (1-\beta_B\cos\theta_{B})\\
    & & \times
    \frac{d\sigma_{AB\to AC}}{d\varepsilon}(\varepsilon^\prime,\varepsilon,\mathbf{p}_B),
    \label{ReinjectionTermB}
\end{eqnarray}
where $d\sigma_{AB\to AC}/d\varepsilon (\varepsilon,\varepsilon',\mathbf{p}_B)$ is the differential cross section for $AB \to AC$ with $A$ having an initial energy $\varepsilon'$ and final energy $\varepsilon$. Finally, we calculate $\mathcal{C}$ via
\begin{eqnarray}\nonumber
    \mathcal{C}_{B\to A}=c \sum_{B,C,D}& & \int d^3\mathbf{p}_C \frac{dn_C^{{\rm C}\nu{\rm B}}}{d^3\mathbf{p}_C}(1-\beta_C\cos\theta_C)\\ & & \;\;\times
    \frac{d\sigma_{BC\to AD}}{d\varepsilon}(\varepsilon',\varepsilon,\mathbf{p}_C).
    \label{ReinjectionTermC}
\end{eqnarray}

For the calculation of the cross sections $\sigma_{AB\to CD}$ (for the process $A+B\to C+D$) and differential cross sections, we consider an incident (ultrarelativistic) neutrino with energy $\varepsilon_A$ propagating in the $+\hat{x}_3$ direction, while the target neutrino has energy $\varepsilon_B$. The angle $\theta_B$ is the angle between the incident and target momenta. The C$\nu$B is isotropic, so we can orient the axes such that $\mathbf{p}_A$ and $\mathbf{p}_B$ lie in the $x_1-x_3$ plane. The outgoing particles are ultrarelativistic.

One of the outgoing neutrinos, with energy $\varepsilon_C$, makes an angle $\theta_C$ with the incident neutrino. These quantities are related by
\begin{equation}
\varepsilon_C = \frac{\varepsilon_A\varepsilon_B(1-\beta_B\cos\theta_{B})}{\varepsilon_A(1-\cos\theta_{C})+\varepsilon_B(1-\beta_B\cos\theta_{BC})},
\label{ScatteringAngleFormula}
\end{equation}
where $\cos\theta_{BC} = \cos\theta_B\cos\theta_C+\sin\theta_B\sin\theta_C\cos\phi_C$ and $\phi_C$ is the azimuthal angle for $\mathbf{p}_C$. In our case, since $ \varepsilon_A/\varepsilon_B\gtrsim 10^{18}$, even at high redshifts, the second term in the denominator is only relevant when $\theta_C\lesssim 10^{-9}$. In such a limit, we have $\varepsilon_B(1-\beta_B\cos\theta_{BC})\to \varepsilon_B(1-\beta_B + \mathcal{O}(\theta_C))$. Therefore, any effects that $\phi_C$ may have on $\varepsilon_C$ can be neglected. 

We then define the Mandelstam variables
\begin{eqnarray}
s = 2\varepsilon_A\varepsilon_B(1-\beta_B\cos\theta_B)\\
t = -2\varepsilon_A\varepsilon_C(1-\cos\theta_C),
\end{eqnarray}
where neutrinos $A$ and $C$ are ultrarelativistic.

The calculation of Eq.~\ref{AbsorptionTermA} only requires the total cross sections $\sigma_{AB}(s)$, which can be found in Ref.~\cite{PhysRevD.47.5247}. 
We use the differential cross sections for the integrals in Eqs.~\ref{ReinjectionTermB} and \ref{ReinjectionTermC}. For convenience, we use the Lorentz-invariant differential cross section $d\sigma/dt$ in terms of the Mandelstam variables $s$ and $t$
\begin{equation}
\frac{d\sigma}{dt} = \frac{1}{16\pi s^2} \langle|\mathcal{M}|\rangle^2,
\end{equation}
where $\langle|\mathcal{M}^2|\rangle$ is the Lorentz-invariant amplitude, averaged over C$\nu$B spins. Each process has its own expression for the amplitude:
\begin{widetext}
\noindent
for $\bar{\nu}_i \nu_j \to \bar{\nu}_i \nu_j$:
\begin{equation}
\langle|\mathcal{M}|^2\rangle = 8G_F^2m_Z^4\left[\dfrac{(s+t)^2}{(t-m_Z^2)^2}+\delta_{ij}\left\{\dfrac{(s+t)^2}{(s-m_Z^2)^2+m_Z^2\Gamma_Z^2}+ 2\dfrac{(s-m_Z^2)(s+t)^2}{(t-m_Z^2)((s-m_Z^2)^2+m_Z^2\Gamma_Z^2)}\right\}\right]
\end{equation}
\noindent
for $\nu_i\nu_j\to\nu_i\nu_j$:
\begin{equation}
\langle|\mathcal{M}|^2\rangle = 8G_F^2m_Z^4s^2\left[\dfrac{1}{(t-m_Z^2)^2}+\delta_{ij}\left\{\dfrac{1}{(s+t+m_Z^2)^2}
 - \dfrac{2}{(t-m_Z^2)(s+t+m_Z^2)}\right\}\right]
\end{equation}
\end{widetext}

The amplitude for $\bar{\nu}_i \nu_j \to \bar{\nu}_i \nu_j$ ($\nu_i\nu_j\to\nu_i\nu_j$) is the same as the amplitude for $\nu_i \bar{\nu}_j \to \nu_i \bar{\nu}_j$ ($\bar{\nu}_i\bar{\nu}_j\to\bar{\nu}_i
\bar{\nu}_j$). 

Regarding the discretization of the energy bins and the target neutrino phase space, we proceed as follows:
\begin{enumerate}
    \item The energy range $[10^{8}\;{\rm GeV}, 10^{17}\;{\rm GeV}]$ is discretized in $\log_{10}\varepsilon$, with 20 bins per energy decade. For the C$\nu$B, the absolute momentum $|\mathbf{p}|$ is discretized in $\log_{10}|\mathbf{p}|$, with 20 bins per momentum decade in the range $[10^{-5}\;{\rm eV},10\;{\rm eV}]$. 
    \item To account for the energy dependence of the cross sections within an energy bin $i$ with energy $\varepsilon_i$, we define the energies $\varepsilon_{i\pm 1/2}=(\varepsilon_i+\varepsilon_{i\pm 1})/2$. We then average the interaction terms over the energy range $[\varepsilon_{i-1/2},\varepsilon_{i+1/2}$]. This averaging procedure is particularly useful when $\varepsilon_i$ is close to the resonance energies when considering resonant interactions.
\end{enumerate}

We now write the interaction rate at energy $\varepsilon_i$, $\mathcal{A}_{A,i}$ as follows. Starting from Eq.~\ref{AbsorptionTermA}, we first convert $d^3\mathbf{p}_B = 2\pi |\mathbf{p}_B|^2  d|\mathbf{p}_B|d\cos\theta_B $ , where the isotropy of the C$\nu$B is assumed in the integration over the azimuthal angle. 
The integral over $|\mathbf{p}_B|$ is replaced with a discrete sum over the C$\nu$B momenta bins $|\mathbf{p}_B|_k$, with the width $\Delta|\mathbf{p}_B|_k=(|\mathbf{p}_B|_{k+1}-|\mathbf{p}_B|_{k-1})/2$. 
We then have
\begin{eqnarray}\nonumber
\mathcal{A}_{A,i} &= &\sum_{B,k} \frac{\Delta|\mathbf{p}_B|_k}{4\pi^2}\frac{|\mathbf{p}_{B}|_k^2}{e^{|\mathbf{p}_B|_k/T(z)}+1}\frac{1}{\Delta\varepsilon_i}\int_{\varepsilon_{i-1/2}}^{\varepsilon_{i+1/2}}d\varepsilon\\
& & \times \int_{-1}^1 d\cos\theta_B(1-\beta_{B,k}\cos\theta_B)\sigma_{AB}(s),
\end{eqnarray}
where $\beta_{B,k}$ is the speed corresponding to the momentum $|\mathbf{p}_B|_k$. Notice that the integral over $\varepsilon$ is used to take an average at each bin $\varepsilon_i$, where it is necessary to make sure that the binning process does not miss the resonance effect. The integral over $\theta_B$ alone may not be enough when C$\nu$B neutrinos are nonrelativistic.
We find that averaging over $\varepsilon$ and integrating over $\theta_B$ are enough to encompass the $Z$-resonance effect properly; including an additional average over the target momenta provides a small improvement at the cost of significantly longer computation times. 
The following integral
\begin{equation}
\int_{-1}^1 d\cos\theta_B(1-\beta_{B,k}\cos\theta_B)\sigma_{AB}(s)
\end{equation}
can be calculated for different values of $\varepsilon$, target C$\nu$B momenta $|\mathbf{p}_B|_k$, and neutrino species $B$. Here, the species $B$ specifies the neutrino mass $m_B$ and the interaction channel between $A$ and $B$. The change of variables through $s = 2\varepsilon\varepsilon_{B,k}(1-\beta_{B,k}\cos\theta_B)$ reduces the integral to
\begin{equation}
I^\mathcal{A}_{AB,ik}=\frac{1}{4\varepsilon_{B,k}^2\beta_{B,k}\Delta\varepsilon_i}\int_{\varepsilon_{i-1/2}}^{\varepsilon_{i+1/2}}\frac{d\varepsilon}{\varepsilon^2}\int_{s_-}^{s_+}ds\;s\sigma_{AB}(s),
\end{equation}
where $s_\pm = 2\varepsilon\varepsilon_{B,k}(1\pm \beta_{B,k})$. With this expression, the average over $\varepsilon_i$ can be precomputed for different $\varepsilon_i$ and $|\mathbf{p}_B|_k$, allowing us to recalculate interaction terms on the fly without compromising computational costs. 
We then have
\begin{equation}
\mathcal{A}_{A,i}=
\sum_{B,\;k} \frac{\Delta|\mathbf{p}_B|_k}{4\pi^2}\frac{|\mathbf{p}_{B}|_k^2}{e^{|\mathbf{p}_B|_k/T(z)}+1}I^\mathcal{A}_{AB,ik}.
\label{DiscretizedA}
\end{equation}
We can proceed similarly for $\mathcal{B}$ and $\mathcal{C}$. 
As in Eq.~\ref{DiscretizedA}, we write
\begin{equation}
\mathcal{B}_{A,j\to i}=
\sum_{B,k} \frac{\Delta|\mathbf{p}_B|_k}{4\pi^2}\frac{|\mathbf{p}_{B}|_k^2}{e^{|\mathbf{p}_B|_k/T(z)}+1}I^\mathcal{B}_{AB,ijk},
\label{DiscretizedB}
\end{equation}
where we define
\begin{eqnarray}\nonumber
I^\mathcal{B}_{AB,ijk} = & & \sum_C \frac{1}{\Delta\varepsilon_i}\int_{\varepsilon_{i-1/2}}^{\varepsilon_{i+1/2}}d\varepsilon
\frac{1}{\Delta\varepsilon_j}\int_{\varepsilon_{j-1/2}}^{\varepsilon_{j+1/2}}d\varepsilon^\prime\\\nonumber
& & \int_{-1}^1d\cos\theta_B (1-\beta_{B,k}\cos\theta_B)\\
& & \times \frac{d\sigma_{AB\to AC}}{d\varepsilon}(\varepsilon^\prime,\varepsilon,|\mathbf{p}_B|_k),
\end{eqnarray}
where an additional variable $\varepsilon^\prime$ is introduced. 
At this point, we can first use $d\sigma/dt$ to perform the energy averages over $\varepsilon$. We then have
\begin{eqnarray}\nonumber
I^\mathcal{B}_{AB,ijk} = & & \sum_C \frac{1}{\Delta\varepsilon_i\Delta\varepsilon_j}
\int_{\varepsilon_{j-1/2}}^{\varepsilon_{j+1/2}}d\varepsilon^\prime\int_{-1}^1d\cos\theta_B\\
& &  \times (1-\beta_{B,k}\cos\theta_B)\int_{t_-}^{t_+} \frac{d\sigma_{AB\to AC}}{dt}dt,
\end{eqnarray}
where $t_\pm = -2\varepsilon^\prime\varepsilon_{i\pm 1/2}(1-\cos\theta_\pm)$. Here, $\cos\theta_\pm$ is calculated from Eq.~\ref{ScatteringAngleFormula}. The advantage of integrating $d\sigma/dt$ instead of $d\sigma/d\varepsilon$ is that the former is easier to perform analytically.

\bibliography{ms}

\end{document}